\documentclass[aps,ams,prc,tightenlines,superscriptaddress,showpacs,floatfix,nofootinbib]{revtex4-1}

\usepackage{epsfig, bm}
\usepackage{amsmath}
\usepackage{multirow}
\usepackage{amsmath}
\usepackage{feynmf}
\usepackage{footmisc}
\usepackage{graphicx}
\usepackage{color}
\usepackage[normalem]{ulem}  

\renewcommand\sout{\bgroup \color{red} \ULdepth=-.5ex \ULset}

\newcommand{\Ex}[2]{\ifmmode{#1\times10^{#2}}\else{$#1\times10^{#2}$}\fi}

\begin{document}
\title{QCD sum rules for the $\Delta$ isobar in neutron matter}

\author{Jesuel Marques L.}
\email[]{jesuel@ift.unesp.br}
\affiliation{Instituto de Física Teórica, Universidade Estadual Paulista, São Paulo 01140-070, Brazil}
\affiliation{Department of Physics
and Institute of Physics and Applied Physics, Yonsei University,
Seoul 03722, Republic of Korea}

\author{Su Houng Lee} 
\email[]{suhoung@yonsei.ac.kr}
\affiliation{Department of Physics
and Institute of Physics and Applied Physics, Yonsei University,
Seoul 03722, Republic of Korea}

\author{Aaron Park} 
\email[]{aaron.park@yonsei.ac.kr}
\affiliation{Department of Physics
and Institute of Physics and Applied Physics, Yonsei University,
Seoul 03722, Republic of Korea}

\author{R. D. Matheus}
\email[]{matheus@ift.unesp.br}
\affiliation{Instituto de Física Teórica, Universidade Estadual Paulista, São Paulo 01140-070, Brazil}

\author{Kie Sang Jeong}
\email[]{kiesang.jeong@apctp.org}
\affiliation{Asia Pacific Center for Theoretical Physics, Pohang, Gyeongbuk 37673, Republic of Korea}

\date{\today}

\begin{abstract}
We study the properties of the $\Delta$ isobar in the symmetric and  asymmetric nuclear matter using the QCD sum rules approach based on the energy dispersion relation.
Allowing for different continuum thresholds for the polarization  tensors with different dimensions, we find stable masses for the $\Delta$ in both the vacuum and the medium.  Compared to the nucleon case, we find that the vector repulsion is smaller for the $\Delta$ while the scalar attraction is similar  (75~MeV vector repulsion and  200~MeV scalar attraction in the symmetric matter).  The smaller vector repulsion can be understood using the Pauli principle and a constituent quark model. Also, the isospin dependence of the quasiparticle energy, which mainly comes from the vector self-energy, is quite weak.  We also allow for an explicit $\pi -N$ continuum contribution to the polarization function but find its effect to be  minimal.  Phenomenological consequences of our results are discussed.
\end{abstract}

\maketitle

\section{Introduction}

The $\Delta$ isobar is the lowest excitation of the nucleon that
has been identified experimentally, which was done more than 70 years ago~\cite{Anderson:1952nw}.  Since then, it has been  a subject of interest in relation to nuclear physics phenomena and in particular to pion dynamics in nuclear matter~\cite{Brown:1975di}.  Recently, there has been a renewed interest in the changes of $\Delta$ properties in  asymmetric nuclear matter, as large  changes will soften the  neutron star equation of state and hence influence  the maximum allowed mass for the neutron star~\cite{Cai:2015hya}. Although the mass shift of the $\Delta$ is expected to be small in nuclear matter ~\cite{Post:2003hu, Korpa:2003bc, Korpa:2008ut}, there are indications from nuclear target experiments that there are non trivial changes in the $\Delta$  properties in nuclear matter ~\cite{Ellegaard:1989gr,Trzaska:1991ww}.  Furthermore, heavy ion data points to larger mass shifts at higher density~\cite{Barrette:1994kq,Hjort:1997wk,Pelte:1999ir,Eskef:2001qg,Bi:2005ca}.

The properties of the $\Delta$ in symmetric nuclear matter were studied in a phenomenological model with nucleon resonances~\cite{Post:2003hu} and with self-consistent calculations considering the covariant particle-hole vertex~\cite{Korpa:2003bc, Korpa:2008ut}, in a chiral quark-meson model~\cite{Christov:1990tc}, in a relativistic chiral effective field theory~\cite{Lee:2007zzm}, and in the QCD sum rule approaches~\cite{Jin:1994vw,Johnson:1995sk, Azizi:2016hbr}.  
In Refs.~\cite{Korpa:2003bc, Korpa:2008ut} in particular, a fully relativistic and self-consistent
many body approach was developed that consistently includes Migdal’s short range correlations effects.  With a realistic parameters, the work predicted a downward shift of about 50 MeV for the $\Delta$ resonance at nuclear saturation density.  Here, we will use the QCD sum rule approach to express the in-medium modification of $\Delta$ state in terms of chiral symmetry restoration and subsequent change of condensate values~\cite{Shifman:1978bx, Ioffe:1981kw, Reinders:1984sr, Drukarev:1988kd, Cohen:1991js, Cohen:1994wm} and improve on the previous works by allowing for different continuum thresholds for the polarization  tensors with different dimensions, and treating  the 4-quark condensates in greater detail.  We then apply this approach to study the $\Delta$ in the asymmetric nuclear matter, where we allow for the difference in the density  of  the protons and neutrons. Additionally, we will investigate the effect of explicitly allowing for the $\pi-N$ continuum contribution to the phenomenological side of the  polarization function as the self-energy of the  $\Delta$ can be significantly affected by the $\pi-N$ state in both  the vacuum and medium~\cite{Korpa:2003bc, Korpa:2008ut}.  We find that while the strength of the scalar attraction is similar to that obtained in the nucleon case, the vector repulsion is smaller, which can be understood using the Pauli principle and a  constituent quark model.  Our result shows that the mass of $\Delta^{++}$ is reduced by 125 MeV (200 MeV scalar attraction and 75 MeV vector repulsion) in symmetric nuclear matter  and by 145 MeV (190 MeV scalar attraction and 45 MeV vector repulsion) in neutron matter  at  nuclear matter density.  

The paper is organized as follows. In Sec.~II, we introduce the currents for the $\Delta$ and discuss the QCD sum rule formalism for this particle. In Sec.~ III, the results of our sum rule analysis are discussed.  In Sec.~IV, we discuss why the vector repulsion for the $\Delta$ should be smaller than that for the nucleon within a constituent quark model.   Discussions and conclusions are given in Sec.~V.

\section{QCD sum rules for the $\Delta$ isobar state}\label{sec2}

\subsection{Correlation of spin-$\frac{3}{2}$ state}
 
In this study, we have calculated in-medium properties of the $\Delta$ isobar state.  As discussed in the literature~\cite{Shifman:1978bx, Azizi:2016hbr, Ioffe:1981kw, Reinders:1984sr, Drukarev:1988kd, Cohen:1991js, Jin:1994vw}, we first had to choose the proper interpolating current that has the same quantum number as the hadron of interest. If the coupling strength between the $\Delta$ isobar state and the  interpolating field is strong enough, information on the physical properties of the $\Delta$ can be obtained from the correlation function:
\begin{align}
\Pi_{\mu\nu}(q) \equiv i \int d^4 x e^{iqx} \langle\Psi_0\vert
\textrm{T}[\eta_{\mu}(x)\bar{\eta}_{\nu}(0)]\vert\Psi_0\rangle,\label{corr}
\end{align}
where  $\vert\Psi_0\rangle$ is the parity and time-reversal symmetric ground state and $\eta_{\mu}(x)$ is an interpolating field for the $\Delta$ isobar state:
 \begin{align}
\langle \Psi_0 \vert \eta_{\mu}(0) \vert\ \Delta (q,s)\rangle= \lambda \psi_{\mu}^{(s)}(q),
\end{align}
where $\lambda$ is the coupling to the $\Delta$ state and $ \psi_{\mu}^{(s)}$ is  the $\Delta$ isobar wave function with spin  index $s$ and momentum $q$. In the phenomeological sense, the wave function  can be regarded as a Rarita-Schwinger (RS) field~\cite{Rarita:1941mf}. In the RS formalism, the Lorentz indices represent the bosonic nature and the Dirac indices represent the fermionic nature of a spin $\frac{2n+1}{2}$ system, where $n$ is an integer. To account for the correct degrees of freedom of the spin-$\frac{3}{2}$ system, the RS field should satisfy the following gauge constraints:
\begin{align}
q^{\mu}\psi^{(s)}_{\mu} (q)  =0,\quad \gamma^{\mu} \psi^{(s)}_{\mu} (q) =0,
\end{align}
where `$q$' represents the momentum of the on-shell $\Delta$ isobar state.

In a vacuum, the propagator of the relativistic spin-$\frac{3}{2}$ field can be obtained as~\cite{Nath:1971wp, Benmerrouche:1989uc}
\begin{align}
S^{3/2}_{\mu\nu}(q) = \frac{1}{\slash \hspace{-0.2cm} q - m_{\Delta}}\left(   g_{\mu\nu} - \frac{1}{3}\gamma_{\mu} \gamma_{\nu}-\frac{2q_{\mu}q_{\nu}}{3q^2}-\frac{1}{3q^2}(\slash \hspace{-0.2cm} q \gamma_\mu q_\nu  + q_\mu \gamma_\nu \slash \hspace{-0.2cm} q) \right).
\end{align}
Making use of the Dirac equation $(\slash \hspace{-0.2cm} q - m_{\Delta})\psi_{\mu}^{(s)} (q)=0$ and  the normalization condition  $\psi^{(s)}_{\mu}  \psi^{(s)\mu} =-m_{\Delta}$ of the wave function, the $\Delta$ isobar state in the correlator can be described as
\begin{align}
\sum_{s} \langle 0 \vert \eta_{\mu}(0) \vert\ \Delta (q,s)\rangle \langle \Delta (q,s) \vert \bar{\eta}_{\nu}(0)\vert 0  \rangle= - \lambda^2(\slash \hspace{-0.2cm} q +m_{\Delta})\left(   g_{\mu\nu} - \frac{1}{3}\gamma_{\mu} \gamma_{\nu}-\frac{2q_{\mu}q_{\nu}}{3q^2}-\frac{1}{3q^2}(\slash \hspace{-0.2cm} q \gamma_\mu q_\nu  + q_\mu \gamma_\nu \slash \hspace{-0.2cm} q) \right).\label{corrvac}
\end{align}
On the other hand, as the interpolating field is not the properly gauged RS field, the correlator~\eqref{corr} has following structure:
\begin{align}
\Pi_{\mu\nu}(q)= \left[\Pi_{0}(q)+ \Pi_{q}(q) \slash \hspace{-0.2cm} q\right]  \mathcal{P}^{3/2} (q) + \left[\Pi^{11}_{0}(q) + \Pi^{11}_{1}(q)  \slash \hspace{-0.2cm} q \right]\mathcal{P}^{1/2}_{11} (q)+ \left[\Pi^{22}_{0}(q) + \Pi^{22}_{1} (q) \slash \hspace{-0.2cm} q \right]\mathcal{P}^{1/2}_{22} (q) +\cdots,
\end{align}
where the spin projections, satisfying the constraint $\mathcal{P}^{3/2}_{ } (q) + \mathcal{P}^{1/2}_{11 } (q)+\mathcal{P}^{1/2}_{22 } (q)=g_{\mu\nu}$, can be summarized as
\begin{align}
\mathcal{P}^{3/2}_{ } (q) &= g_{\mu\nu} - \frac{1}{3}\gamma_{\mu} \gamma_{\nu}-\frac{1}{3q^2}(\slash \hspace{-0.2cm} q \gamma_\mu q_\nu  + q_\mu \gamma_\nu \slash \hspace{-0.2cm} q) ,\\ 
\mathcal{P}^{1/2}_{11 } (q) &= \frac{1}{3}\gamma_{\mu} \gamma_{\nu} -\frac{q_{\mu}q_{\nu}}{q^2} +\frac{1}{3q^2}(\slash \hspace{-0.2cm} q \gamma_\mu q_\nu  + q_\mu \gamma_\nu \slash \hspace{-0.2cm} q),\\
\mathcal{P}^{1/2}_{22 } (q) &= \frac{q_{\mu}q_{\nu}}{q^2}.
\end{align}
Thus, to extract the $\Delta$ properties, one can concentrate on the 
 $g_{\mu\nu}$ tensor structure and extract the $[\Pi_{0}(q) + \Pi_{q}(q) \slash \hspace{-0.2cm} q] $ part of the polarization function that 
has information only about the spin $\frac{3}{2}$ system:
\begin{align}
\Pi_{\mu\nu}(q) =\left[ \Pi_{s}(q)+ \Pi_{q}(q) \slash \hspace{-0.2cm} q \right] g_{\mu\nu}+ \cdots \Rightarrow \frac{\lambda^2}{\slash \hspace{-0.2cm} q - m_{\Delta}}g_{\mu\nu} + \cdots.\label{phenvac1}
\end{align}

\subsubsection{Phenomenological side}

Now, consider taking the medium expectation value of the correlation function 
characterized by the nuclear  density
$\rho=\rho_n+\rho_p$, the matter velocity $u_\mu$, and the iso-spin
asymmetry factor $I=(\rho_n-\rho_p)/(\rho_n+\rho_p)$.  
In the mean-field description of the  quasi-$\Delta$ isobar state in the medium,  the quasi-particle wave function $\tilde{\psi}_{\mu}^{(s)}$ will satisfy the  equation of motion $[\slash \hspace{-0.2cm} q  - m_{\Delta}-\Sigma(q,u)] \tilde{\psi}_{\mu}^{(s)} (q)=0$, where $\Sigma(q,u) = \Sigma_{v}(q^2,qu) \slash \hspace{-0.2cm} u +  \Sigma_{s}(q^2,qu) $. The functions $\Sigma_s$, $\Sigma_v$ are the the scalar and vector self-energies, respectively. Then, the in-medium correlation can be written as
\begin{align} 
\sum_{s} \langle \Psi_0  \vert \eta_{\mu}(0)  \vert\ \Delta (q,s)\rangle \langle \Delta (q,s) \vert \bar{\eta}_{\nu}(0) \vert \Psi_0  \rangle= - {\lambda^{*}}^2 ( \slash \hspace{-0.2cm} q  -\Sigma_{v} \slash \hspace{-0.2cm} u  +m_{\Delta} + \Sigma_{s}) (g_{\mu\nu}+\cdots),
\end{align}
and the corresponding phenomenological structure of the in-medium correlator  is ~\cite{Jin:1994vw}
\begin{align}
\Pi_{\mu\nu}(q)  &= \left(\Pi_{s}(q^2,qu)+ \Pi_{q}(q^2,qu) \slash \hspace{-0.2cm} q  + \Pi_{v}(q^2, qu) \slash \hspace{-0.2cm} u \right) g_{\mu\nu}+ \cdots, \nonumber\\
& \Rightarrow  \frac{ {\lambda^{*}}^2 }{\slash \hspace{-0.2cm} q -\Sigma_{v} \slash \hspace{-0.2cm} u- m_{\Delta}- \Sigma_{s}  }g_{\mu\nu} + \cdots.
\label{corrmed}
\end{align}

\subsubsection{Operator product expansion:} 
Each invariant can be expressed in terms of  the QCD degrees  of freedom via the  OPE in  $ q_0^2\rightarrow \infty, \vert \vec{q} \vert \rightarrow \textrm{fixed} $ limit:
\begin{align}
\Pi_i(q^2,q_0^2) &=\sum_n C^i_n(q^2,q_0^2)\langle
\hat{O}_n\rangle_{\rho,I},
\end{align}
where $C^i_n(q^2,q_0^2)$ represent the Wilson coefficients and  $\langle \hat{O}_n\rangle_{\rho,I}$ represents the in-medium condensate. 

\subsubsection{Dispersion relation}

 The correlation function satisfies the following dispersion relation on the complex
$\omega$ plane:
\begin{align}
\Pi_i(q_0,\vert\vec{q}\vert)&=\frac{1}{2\pi i}\int^\infty_{-\infty}
d\omega  \frac{ \Delta\Pi_i(\omega,\vert\vec{q}\vert)}{\omega-q_0}
+F_n(q_0 ,\vert\vec{q}\vert), \label{dis-eo}
\end{align}
where $F _n(q_0 ,\vert\vec{q}\vert) \equiv
F^e_n(q_0^2,\vert\vec{q}\vert) +q_0F^o_n(q_0^2,\vert\vec{q}\vert)$
is a finite-order polynomial. The discontinuity
$\Delta\Pi_i(\omega,\vert\vec{q}\vert)$ is defined as follows:
\begin{align}
\Delta\Pi_i(\omega,\vert\vec{q}\vert)
&\equiv\lim_{\epsilon\rightarrow0^{+}}
[\Pi_i(\omega+i\epsilon,\vert\vec{q}\vert)-\Pi_i
(\omega-i\epsilon,\vert\vec{q}\vert)]=2i\textrm{Im}[\Pi_i(\omega+i\epsilon,\vert\vec{q}\vert)]
\nonumber\\
&=\Delta\Pi^e_i(\omega^2,\vert\vec{q}\vert)+\omega
\Delta\Pi^o_i(\omega^2,\vert\vec{q}\vert).\label{discm}
\end{align}
All the possible resonances including the ground state are contained in the discontinuity \eqref{discm}. By using
these relations, the invariants can be decomposed into an even and an odd
part in powers of $q_0$,  each part having the following dispersion relation at
fixed $\vert\vec{q}\vert$:
\begin{align}
\Pi_i(q_0,\vert\vec{q}\vert)&=\Pi_i^e(q_0^2,\vert\vec{q}\vert)+q_0
\Pi_i^o(q_0^2,\vert\vec{q}\vert),\\
\Pi_i^e(q_0^2,\vert\vec{q}\vert)& =\frac{1}{2\pi
i}\int^\infty_{-\infty} d\omega \frac{\omega^2
}{\omega^2-q_0^2}\Delta\Pi^o_i(\omega^2,\vert\vec{q}\vert)
+F^e_n(q_0^2,\vert\vec{q}\vert),\label{dise}
  \\
\Pi_i^o(q_0^2,\vert\vec{q}\vert)& =\frac{1}{2\pi
i}\int^\infty_{-\infty} d\omega \frac{
 1}{\omega^2-q_0^2}\Delta\Pi^e_i(\omega^2,\vert\vec{q}\vert)
+F^o_n(q_0^2,\vert\vec{q}\vert),\label{diso}
\end{align}
where $\Delta\Pi_i^e(q_0^2,\vert\vec{q}\vert)$ and
$\Delta\Pi_i^o(q_0^2,\vert\vec{q}\vert)$ are even functions of
$q_0$. In the vacuum limit ($u_{\mu} \rightarrow 0$ and $\rho\rightarrow 0 $), $\Pi_i^e(q_0^2,\vert\vec{q}\vert) $ reduces to
\begin{align}
\Pi_i^e(q_0^2,\vert\vec{q}\vert) \Rightarrow \Pi_i (s) =\frac{1}{ \pi
i}\int^\infty_{0} ds \frac{ \Delta\Pi_i(s)
}{s-q^2}+F_n(s),
\end{align}
and  $\Pi_i^o(q_0^2,\vert\vec{q}\vert) $ vanishes.

\subsection{Currents for the $\Delta$ isobar state}

In the $\mbox{SU(2)}$ limit, one can represent low-lying baryon states by a direct product of the light quarks~\cite{Cohen:1996sb}: 
\begin{align}
\left[ \left( \frac{1}{2},0 \right) \oplus \left(0,  \frac{1}{2}\right)\right]^3& = \left[ \left( \frac{3}{2},0 \right)\oplus\left(0,  \frac{3}{2}\right)\right]+3\left[ \left( \frac{1}{2},1 \right)\oplus\left(1,  \frac{1}{2}\right)\right]+\left[ \left( I=\frac{1}{2} \right)~\textrm{representations} \right],
\end{align}
where we have used the notation for chiral multiplets, with the first and second
numbers in the parenthesis refering to $\mbox{SU(2)}_L$ and $\mbox{SU(2)}_R$ representations, respectively.
 
As the $\Delta$ isobar has the quantum numbers of spin-$\frac{3}{2}$ and isospin-$\frac{3}{2}$, it can be described in either the $  \left( \frac{3}{2},0 \right) \oplus \left(0,  \frac{3}{2}\right) $ or the $  \left( \frac{1}{2},1 \right) \oplus \left(1,  \frac{1}{2}\right) $ representations.

For the $\Delta^{++}$ state, one  takes all the quarks to be the $u$ quarks.  Then, in the $\left(\frac{3}{2},0 \right) \oplus \left(0,\frac{3}{2} \right)$ representation, a possible   interpolating current can be chosen as
\begin{align}
\eta^{\Delta}_{\mu\nu}&  \equiv (u^T C \sigma^{\alpha\beta} u) \sigma_{\alpha\beta} \sigma_{\mu\nu} u,\label{tcrnt}
\end{align}
where the antisymmetrized color indices are implied. The $\Delta^{++}$ state can be obtained in the parity-even mode in the correlation function. In the OPE of the correlation function of Eq.~\eqref{tcrnt}, the leading quark condensate contribution appears as a four quark condensate with ${\alpha_s}^2$ correction. Subsequent sum rule analysis may contain non-negligible ambiguity as higher order quark condensates are not well understood at present.

In the $\left( \frac{1}{2}, 1 \right) \oplus \left(1,  \frac{1}{2} \right) $ representation, one possible interpolating field is

\begin{align}  
\eta^{(1)}_{\mu}&\equiv   (u^T C \sigma^{\alpha\beta} u)  \sigma_{\alpha\beta}  \gamma_{\mu}u =4\eta^{(2)}_{\mu} = 4(u^T C \gamma_{\mu} u) u.
\end{align}

The $\eta^{(1)}_{\mu}$ current is renormalization covariant and can be shown to be equivalent up to a numerical factor to $\eta^{(2)}_{\mu}$, introduced by Ioffe~\cite{Ioffe:1981kw}.  In this representation, the chiral condensate appears as the leading quark condensate contribution in the correlation function. Also, as discussed in Ref.~\cite{Ioffe:1981kw}, the spin state of individual quark in $\eta^{(2)}_{\mu} = (u^T C \gamma_{\mu} u) u$ can be understood in a consistent way within the constituent quark model picture. In this work, we concentrate on this 
$\left( \frac{1}{2} ,1 \right) \oplus \left(1, \frac{1}{2} \right) $ representation and use the $\eta^{(2)}_{\mu} = \epsilon_{abc}(u_a^T (x) C \gamma_{\mu} u_b(x) ) u_c(x) $ current that possess the aforementioned benefits. 

\subsection{Operator product expansion and Borel sum rules} 

We will follow the in-medium light quark and gluon condensates as described in Refs.~\cite{Cohen:1994wm, Jeong:2012pa, Jeong:2016qlk}. Here we summarize the density dependences of the condensates. The condensates in linear density approximation can be expressed  as follows:
\begin{align}
\langle\hat{O}\rangle_{\rho,I}=&~\langle\hat{O}\rangle_{\textrm{vac}}+\langle
n\vert\hat{O}\vert n\rangle \rho_n+\langle p \vert \hat{O} \vert p
\rangle\rho_p \nonumber \\
=&~\langle\hat{O}\rangle_{\textrm{vac}}+\frac{1}{2}  (\langle
n\vert\hat{O}\vert n\rangle +\langle p \vert \hat{O} \vert p \rangle
)\rho  +\frac{1}{2} (\langle n\vert\hat{O}\vert n\rangle -\langle p
\vert \hat{O} \vert p \rangle) I \rho, \label{condensatea}
\end{align}
where $\rho=\rho_n+\rho_p$ and $I \rho=\rho_n-\rho_p$.  Also,  $\rho=\rho_0=
0.16~\textrm{fm}^{-3}=(110~\textrm{MeV})^3$ is assigned for the normal nuclear density. Consider an operator $\hat{O}_{u,d}$ composed of either up or down
quarks, respectively.  Using the isospin symmetry relation,
\begin{align}
\langle n\vert\hat{O}_{u,d}\vert n\rangle=\langle
p\vert\hat{O}_{d,u}\vert p\rangle,
\end{align}
the neutron expectation value can be converted into the proton
expectation value. 

The
two-quark operators, Eq.~\eqref{condensatea}, can be arranged  as follows:
\begin{align}
\langle\hat{O}_{u,d}\rangle_{\rho,I}
=\langle\hat{O}_{u,d}\rangle_{\textrm{vac}}+(\langle
p\vert\hat{O}_0\vert p\rangle\mp\langle p \vert \hat{O}_1 \vert p
\rangle I)\rho.\label{condensateb}
\end{align}
Here,  `$\mp$' respectively stands for the $u$ and $d$ quark flavors and the isospin operators are defined as
\begin{align}
\hat{O}_0\equiv\frac{1}{2}(\hat{O}_u+\hat{O}_d),\quad
\hat{O}_1\equiv\frac{1}{2}(\hat{O}_u-\hat{O}_d).\label{isospincom}
\end{align}
All the expectation values will be expressed in terms of the proton
counterparts and be denoted as $\langle p \vert \hat{O} \vert p
\rangle\rightarrow\langle\hat{O}\rangle_p$.

The four quark condensates are denoted  following the notation of Ref.~\cite{Jeong:2016qlk}
\begin{align}
\epsilon_{abc}\epsilon_{a'b'c} \langle \bar{q}_{a'} \Gamma_m^\alpha
q_{a} \bar{q}_{b'} \Gamma_m^{\beta}
q_{b}  \rangle_{\rho,I}=&~\frac{g^{\alpha\beta} }{4} \langle
\bar{q}_{} \Gamma_m  q_{} \bar{q}_{} \Gamma_m
q_{}  \rangle_{\textrm{tr.}}+\left(u^{\alpha}u^\beta
-\frac{g^{\alpha\beta}}{4}\right) \left\langle \bar{q}_{} \Gamma_m q_{}
\bar{q}_{} \Gamma_m
q_{} \right\rangle_{\textrm{s.t.}},\\
\langle \bar{q}_{} \Gamma_m q_{} \bar{q}_{} \Gamma_m
q_{}\rangle_{\textrm{tr.}} =&~\frac{2}{3}\langle\bar{q}
\Gamma_m^\alpha q \bar{q} \Gamma_{m\alpha}
q\rangle_{\textrm{vac}}-2\langle\bar{q}  \Gamma_m^\alpha t^A q
\bar{q} \Gamma_{m\alpha} t^A
q\rangle_{\textrm{vac}}\nonumber\\
& +\sum_{i=\{n,p\}}\left(\frac{2}{3}\langle\bar{q} \Gamma_m^\alpha
q \bar{q} \Gamma_{m\alpha} q\rangle_{i}-2\langle\bar{q}
\Gamma_m^\alpha t^A q_1 \bar{q} \Gamma_{m\alpha} t^A
q\rangle_{i}\right)\rho_i,\\
\langle \bar{q} \Gamma_m q \bar{q} \Gamma_m
q\rangle_{ \textrm{s.t.}}
=&~\sum_{i=\{n,p\}}\left(\frac{2}{3}\langle\bar{q} \Gamma_m q
\bar{q} \Gamma_m q\rangle_{i, \textrm{s.t.}} -2\langle\bar{q}
\Gamma_m t^A q \bar{q} \Gamma_n t^A
q \rangle_{i,\textrm{s.t.}}\right)\rho_i,
\end{align}
where $\Gamma_m=\{I,\gamma_5, \gamma, \gamma_5\gamma, \sigma \}$,  and the 
subscripts $ \textrm{vac} $, $ i $, and $ \textrm{s.t.} $ represent
the vacuum expectation value, nucleon expectation value, and symmetric
traceless matrix elements, respectively. Twist-4 matrix elements are assigned by following estimates as in Refs.~\cite{Choi:1993cu, Jeong:2012pa, Jeong:2016qlk}. The spin-0 and spin-1 condensates values are estimated by using the factorization hypothesis:
\begin{align}
\langle q^a_\alpha\bar{q}^b_\beta
q^c_\gamma\bar{q}^d_\delta\rangle_{\rho,I}&\simeq\langle
q^a_\alpha\bar{q}^b_\beta \rangle_{\rho,I}\langle
q^c_\gamma\bar{q}^d_\delta \rangle_{\rho,I}- \langle
q^a_\alpha\bar{q}^d_\delta \rangle_{\rho,I} \langle
q^c_\gamma\bar{q}^b_\beta \rangle_{\rho,I},\label{factorization}\\
\langle[\bar{u}u] \rangle^2_{\rho,I } & \Rightarrow
k_1\langle\bar{q}q\rangle^2_{\textrm{vac}} +
2f_1\left(\langle[\bar{u}u]_0 \rangle_p  - \langle \bar{u}u_1
\rangle_p I\right)\langle\bar{q}q\rangle_{\textrm{vac}}
\rho,\label{4qscalar}\\
\langle u^\dagger u \rangle_{\rho,I} \langle \bar{u}u\rangle_{\rho,I}&
\Rightarrow k_2  \left(\langle[u^\dagger u]_0 \rangle_p - \langle[u^\dagger u]_1
\rangle_p I\right) \langle\bar{u}u\rangle_{\textrm{vac}}\rho,\label{4qscalar1} 
\end{align}
where the parameters $k_1$, $k_2$ determine the factorization strength for the vacuum piece and $f_1$ determine the medium dependence of the scalar four-quark
condensate. Both  $k_1$ and $k_2$ are set to 0.7 to avoid an overly strong  contribution of four quark condensates which may amplify ambiguity of the  condensates.  Furthermore, this choice leads a good Borel curve that correctly reproduces the mass of the $\Delta$ in the vacuum.
 $f_1=0.1$ has been assigned according to
previous studies, where $ f_1 $ should be weak
($\vert f_1 \vert \ll 1$)
 \cite{Cohen:1994wm, Jeong:2012pa, Jeong:2016qlk}. 

The correlation function~\eqref{corr} contains all possible
states that overlap with the quantum numbers of the interpolating
field as discussed before. Because our interest lies in the self-energies on the quasiparticle
pole, the other excitations should be suppressed. Borel sum rules
can be used for this purpose: The weight function $W(\omega)=
(\omega-\bar{E}_q)e^{-\omega^2/M^2}$ has been applied to the
discontinuity in the dispersion relation and the
corresponding differential operator $\bar{\mathcal{B}}$ has been
applied to the OPE side. Each transformed part will be denoted as
$\overline{\mathcal{W}}_M[\Pi(q_0^2,\vert\vec{q}\vert)]$ and
$\bar{\mathcal{B}}[\Pi(q_0^2,\vert\vec{q}\vert)]$ respectively.
Details for the weighting scheme and the corresponding differential operation in the Borel sum rules are briefly summarized in Appendix~\ref{appenb}.

Borel transformed invariants contain the quasi-antipole $\bar{E}_q$
as an input parameter. As we are following relativistic mean field type phenomenology, the antipole $\bar{E}_q$ is defined
regardless of the  clear pole like structure in the medium.  The exact
value can be determined by solving the self-consistent dispersion
relation:
\begin{align}
\bar{E}_q=\Sigma_v(\bar{E}_q)-\sqrt{\vec{q}^2+m_{\Delta}^{*}(\bar{E}_q)^{2}},\label{qhd}
\end{align}
where $m_{\Delta}^{*}=m_{\Delta}+\Sigma_s$ and $\vert\vec{q}\vert=0~\textrm{MeV}$ will be used, as the quasi-$\Delta$ state is not expected to have its own Fermi level at the normal nuclear density $\rho_0$.

The OPE of each invariant has been calculated as follows:
\begin{align}
\Pi^e_{\Delta,s}(q_0^2,\vert\vec{q}\vert) & = 
-\frac{1}{3\pi^2}q^2\ln(-q^2)\langle
\bar{u}u \rangle_{\rho,I} \\
\Pi^o_{\Delta,s}(q_0^2,\vert\vec{q}\vert) &= \frac{2}{q^2} \langle \bar{u}  \gamma_0  u \bar{u}  u \rangle_{\rho,I},\\
\Pi^e_{\Delta,q}(q_0^2,\vert\vec{q}\vert) & =  \frac{1}{160\pi^4}(q^2)^2\ln(-q^2) +\frac{1}{9\pi^2} \ln(-q^2) \langle u^\dagger i D_0 u  \rangle_{\rho,I}   +\frac{4 }{9\pi^2}\frac{q_0^2}{q^2} \langle u^\dagger i D_0 u  \rangle_{\rho,I} \nonumber \\
&\quad-\frac{5}
{288\pi^2}\ln(-q^2)\left\langle\frac{\alpha_s}{\pi}G^2\right\rangle_{\rho,I}+\frac{1}
{36\pi^2}\ln(-q^2)\left\langle\frac{\alpha_s}{\pi}[(u\cdot G)^2+(u\cdot \tilde{G})^2]\right\rangle_{\rho,I}
\nonumber\\
&\quad+\frac{3}{4q^2}\langle \bar{u}  u \bar{u}  u \rangle -\frac{3}{4 q^2}\langle \bar{u} \gamma_5  u \bar{u} \gamma_5  u \rangle +\frac{5}{4 q^2}\langle \bar{u} \gamma u \bar{u}\gamma u \rangle_\textrm{tr.} -\frac{5}{4 q^2}\langle \bar{u} \gamma_5 \gamma u \bar{u} \gamma_5 \gamma u \rangle_\textrm{tr.}\nonumber\\
&\quad+\frac{1}{8 q^2}\langle \bar{u} \gamma u \bar{u}\gamma u \rangle_\textrm{s.t.} -\frac{9}{8 q^2}\langle \bar{u} \gamma_5 \gamma u \bar{u} \gamma_5 \gamma u \rangle_\textrm{s.t.}-\frac{3}{4 q^2} \langle \bar{u}\sigma u \bar{u} \sigma u \rangle_\textrm{s.t.} \\
\Pi_{\Delta,q}^o(q_0^2,\vert\vec{q}\vert) &= -
\frac{1}{6\pi^2}\ln(-q^2)\langle
u^\dagger u\rangle_{\rho,I},\\
\Pi_{\Delta,u}^e(q_0^2,\vert\vec{q}\vert)& = -\frac{1}{4\pi^2}q^2\ln(-q^2)\langle
u^\dagger u \rangle_{\rho,I}\\
\Pi_{\Delta,u}^o(q_0^2,\vert\vec{q}\vert)& =
\frac{8}{9\pi^2} \ln(-q^2) \langle u^\dagger i D_0 u  \rangle_{\rho,I}  -\frac{1}
{72\pi^2}\ln(-q^2)\left\langle\frac{\alpha_s}{\pi}[(u\cdot G)^2+(u\cdot \tilde{G})^2]\right\rangle_{\rho,I}\nonumber \\
&\quad+\frac{1}{q^2}  \langle \bar{u} \gamma u \bar{u}\gamma u \rangle_\textrm{s.t.} +\frac{1}{q^2}\langle \bar{u} \gamma_5 \gamma u \bar{u} \gamma_5 \gamma u \rangle_\textrm{s.t.},
\end{align}
where the covariant derivative expansion is truncated at first order. Weighted invariants can be summarized as
\begin{align}
\overline{\mathcal{W}}_M^{\textrm{subt.}}[\Pi_{\Delta,s}(q_0^2,\vert\vec{q}\vert)]
&=  \bar{\mathcal{B}}[\Pi^e_{\Delta,s}(q_0^2,\vert\vec{q}\vert)]_{\textrm{subt.}}-\bar{E}_{\Delta}\bar{\mathcal{B}}[\Pi^o_{\Delta,s}(q_0^2,\vert\vec{q}\vert)]_{\textrm{subt.}}\label{opes}\nonumber \\
&=-\frac{1}{3\pi^2}(M^2)^2\langle \bar{u}u \rangle_{\rho,I} \tilde{E}_1 L^{\frac{16}{27}}+2\bar{E}_{\Delta,q} \langle u^\dagger u \rangle_{\rho,I} \langle \bar{u}u \rangle_{\textrm{vac}} L^{\frac{4}{27}},\\
\overline{\mathcal{W}}_M^{\textrm{subt.}}[\Pi_{\Delta,q}(q_0^2,\vert\vec{q}\vert)] &= \bar{\mathcal{B}}[\Pi^e_{\Delta,q}(q_0^2,\vert\vec{q}\vert)]_{\textrm{subt.}}-\bar{E}_{\Delta}\bar{\mathcal{B}}[\Pi^o_{\Delta,q}(q_0^2,\vert\vec{q}\vert)]_{\textrm{subt.}} \nonumber\\
&= -\frac{
1}{80\pi^4}(M^2)^3 \tilde{E}_2 L^{\frac{4}{27}}
 +\frac{1}{9 \pi^2} M^2\tilde{E}_0 \langle u^\dagger i D_0 u  \rangle_{\rho,I}   L^{\frac{4}{27}}-\frac{4}{9 \pi^2} \vec{q}^2 \langle u^\dagger i D_0 u  \rangle_{\rho,I}   L^{\frac{4}{27}}\nonumber\\
 &\quad
+\frac{5}
{288\pi^2} M^2 \tilde{E}_0 \left\langle\frac{\alpha_s}{\pi}G^2\right\rangle_{\rho,I} L^{\frac{4}{27}} -\frac{1}
{36\pi^2} M^2\tilde{E}_0 \left\langle\frac{\alpha_s}{\pi}[(u\cdot G)^2+(u\cdot \tilde{G})^2]\right\rangle_{\rho,I} L^{\frac{4}{27}}\nonumber\\
&\quad-\frac{3}{4}\langle \bar{u}  u \bar{u}  u \rangle L^{\frac{4}{27}} +\frac{3}{4 }\langle \bar{u} \gamma_5  u \bar{u} \gamma_5  u \rangle L^{\frac{4}{27}} -\frac{5}{4 }\langle \bar{u} \gamma u \bar{u}\gamma u \rangle_\textrm{tr.} L^{\frac{4}{27}} +\frac{5}{4 }\langle \bar{u} \gamma_5 \gamma u \bar{u} \gamma_5 \gamma u \rangle_\textrm{tr.} L^{\frac{4}{27}}\nonumber\\
&\quad-\frac{1}{8}\langle \bar{u} \gamma u \bar{u}\gamma u \rangle_\textrm{s.t.} L^{\frac{4}{27}} +\frac{9}{8}\langle \bar{u} \gamma_5 \gamma u \bar{u} \gamma_5 \gamma u \rangle_\textrm{s.t.} L^{\frac{4}{27}}+\frac{3}{4} \langle \bar{u}\sigma u \bar{u} \sigma u \rangle_\textrm{s.t.}  L^{\frac{4}{27}}\nonumber\\
&\quad+ \frac{\bar{E}_{\Delta}}{6\pi^2} M^2 \tilde{E}_0 \langle u^\dagger u \rangle_{\rho,I}L^{\frac{4}{27}},\\
\overline{\mathcal{W}}_M^{\textrm{subt.}}[\Pi_{\Delta,u}(q_0^2,\vert\vec{q}\vert)]& = \bar{\mathcal{B}}[\Pi^e_{\Delta,u}(q_0^2,\vert\vec{q}\vert)]_{\textrm{subt.}}-\bar{E}_{\Delta}\bar{\mathcal{B}}[\Pi^o_{\Delta,u}(q_0^2,\vert\vec{q}\vert)]_{\textrm{subt.}}\nonumber\\
&=\frac{1}{4\pi^2}(M^2)^2\tilde{E}_1 \langle
u^\dagger u \rangle_{\rho,I}L^{\frac{4}{27}}\nonumber\\
&\quad+\bar{E}_{\Delta}\bigg[
\frac{8}{9\pi^2} M^2 \tilde{E}_0 \langle u^\dagger i D_0 u  \rangle_{\rho,I} L^{\frac{4}{27}}-\frac{1}
{72\pi^2} M^2 \tilde{E}_0 \left\langle\frac{\alpha_s}{\pi}[(u\cdot G)^2+(u\cdot \tilde{G})^2]\right\rangle_{\rho,I}L^{\frac{4}{27}}\nonumber\\
&\qquad\quad\quad~ - \langle \bar{u} \gamma u \bar{u}\gamma u \rangle_\textrm{s.t.}L^{\frac{4}{27}} -\langle \bar{u} \gamma_5 \gamma u \bar{u} \gamma_5 \gamma u \rangle_\textrm{s.t.}L^{\frac{4}{27}}\bigg],
\end{align}
where $M$ is the Borel mass. Our OPE differs slightly from the one obtained in a previous study~\cite{Jin:1994vw}. For the scalar invariant, the numeric coefficient of the  $\langle \bar{u}  \gamma_0  u \bar{u}  u \rangle_{\rho,I}$ condensate we obtain is $2$, whereas Jin obtains $4/3$. The anomalous dimensional running corrections are included as
\begin{align}
L^{-2\Gamma_\eta+\Gamma_{O_n}}\equiv\left[\frac{\ln(M/\Lambda_{\textrm{QCD}})}{\ln(\mu/\Lambda_{\textrm{QCD}})}\right]^{-2\Gamma_\eta+\Gamma_{O_n}},
\end{align}
where $\Gamma_\eta$ ($\Gamma_{O_n}$) is the anomalous dimension of
the interpolating current $\eta$  ($\hat{O}_n$), and  $\mu$ is the
separation scale of the OPE taken to be $\mu\simeq 1~\textrm{GeV}$.  The detailed argument for $\Gamma_\eta =-\frac{2}{9}$ came from symmetric quark configuration  and can be found in Ref.~\cite{Peskin:1979mn}.
The OPE continuum effect above ground state has been subtracted by
multiplying the corresponding $\tilde{E}_n$ to all $(M^2)^{n+1}$ terms in
$\overline{\mathcal{W}}_M[\Pi_{\Delta,i}(q_0^2,\vert\vec{q}\vert)]
$ \cite{ Cohen:1994wm}:
\begin{align}
\tilde{E}_0&\equiv1-e^{-s_0^{*}/M^2},\label{contsub0}\\
\tilde{E}_1&\equiv1-e^{-s_0^{*}/M^2}\left(s_0^{*}/M^2+1\right),\label{contsub1}\\
\tilde{E}_2&\equiv1-e^{-s_0^{*}/M^2}\left(s_0^{*2}/2M^4+s_0^{*}/M^2+1\right),\label{contsub2}
\end{align}
where $s_0^{*}\equiv\omega_0^{2}-\vec{q}^2$ and $\omega_0$ is the
energy at the continuum threshold, as briefly explained in Appendix \ref{appenb}. The continuum subtracted invariants have been denoted as $\overline{\mathcal{W}}_M^{\textrm{subt.}}[\Pi_{\Delta,i}(q_0^2,\vert\vec{q}\vert)]$.

\section{Sum rule analysis}\label{sec3}

\subsection{Sum rule analysis for the quasi-$\Delta$ isobar state}

\begin{figure}
\includegraphics[width=6.0cm]{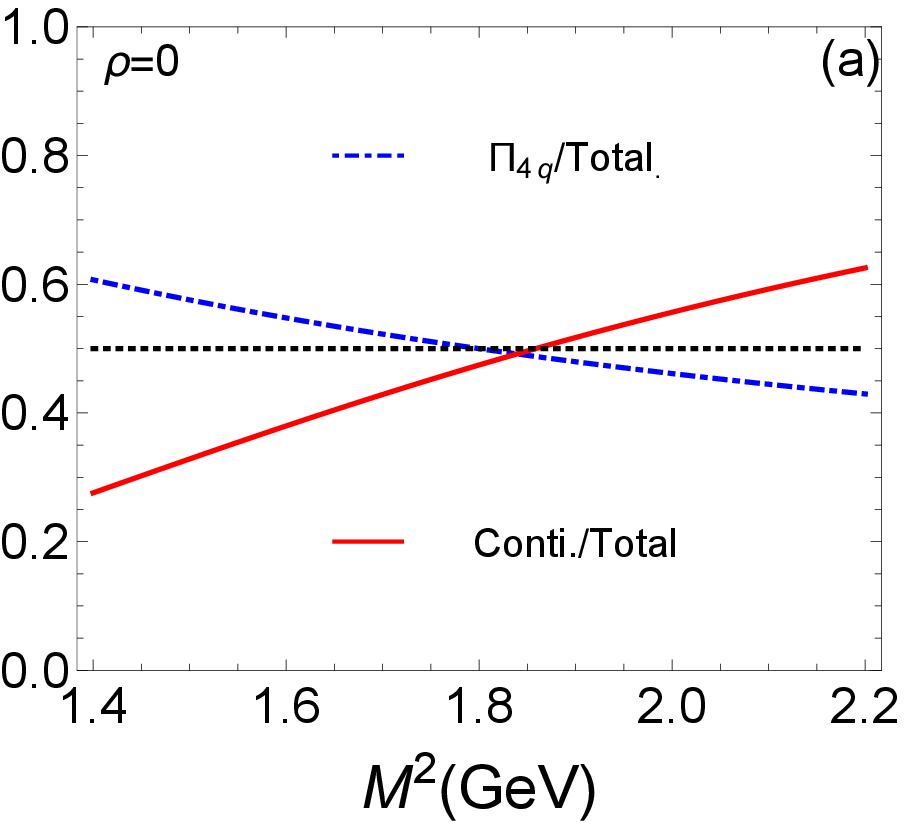} \quad
\includegraphics[width=6.0cm]{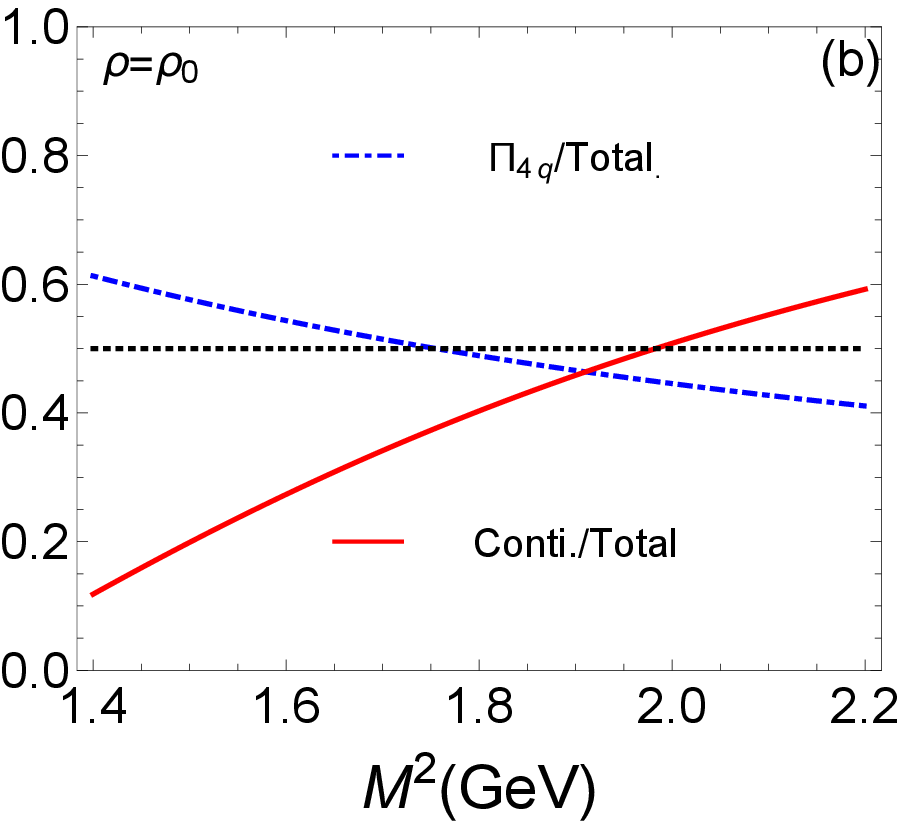} 
\caption{Borel window estimated from  $\overline{\mathcal{W}}_M^{\textrm{subt.}}[\Pi_{\Delta,q}(q_0^2,\vert\vec{q}\vert)] $ (a) in vacuum and (b) in medium. Black dotted line represents 50\%.}\label{fig1}
\end{figure}

If one could calculate the correlation function exactly, the sum rules should not depend on the Borel mass. However, as the OPE is truncated at a finite order, one should find the proper range of Borel mass. The upper bound can be constrained by requiring the OPE continuum  not to exceed $50\%$ of the total OPE contribution while the lower bound can be constrained by the condition that the contribution of four-quark condensates does not exceed $50\%$ of the OPE contribution.
We constrain the Borel window through the weighted invariant $\overline{\mathcal{W}}_M^{\textrm{subt.}}[\Pi_{\Delta,q}(q_0^2,\vert\vec{q}\vert)]$ because this weighted invariant  contains all the contribution of OPE diagrams. In Fig.~\ref{fig1}, one can find that the proper Borel window is very narrow. As the criteria are satisfied near $M^2 \simeq 1.8 ~\textrm{GeV}^2$, the sum rules will be analyzed in the range of $ 1.5~\textrm{GeV}^2 \leq M^2 \leq 2.0~\textrm{GeV}^2$. 

\begin{figure}
\includegraphics[width=6.3cm]{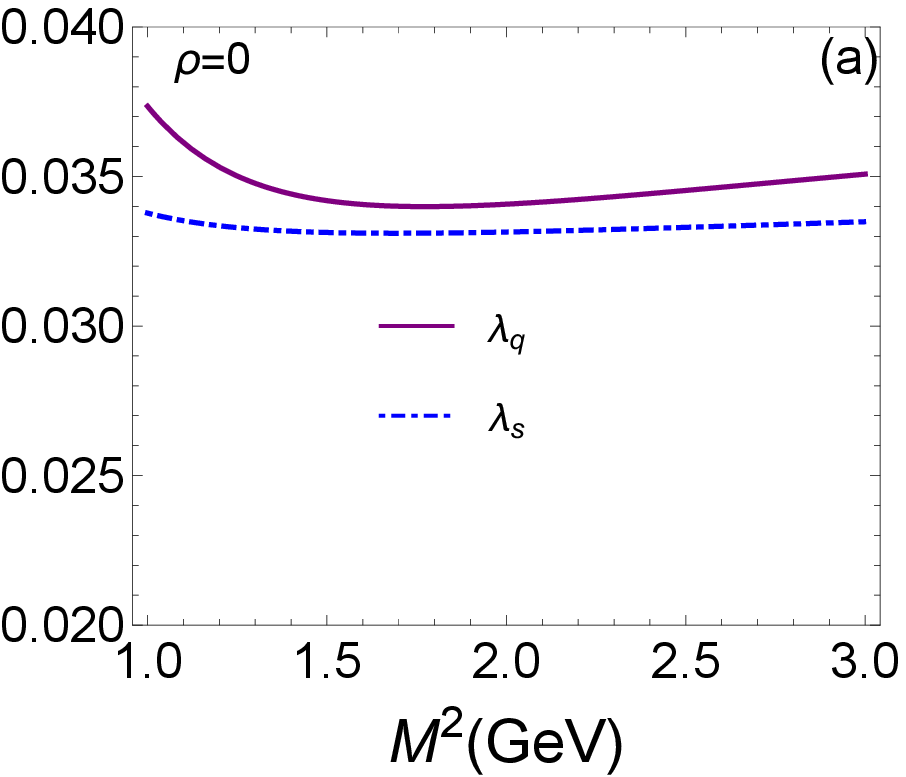} \quad
\includegraphics[width=6.3cm]{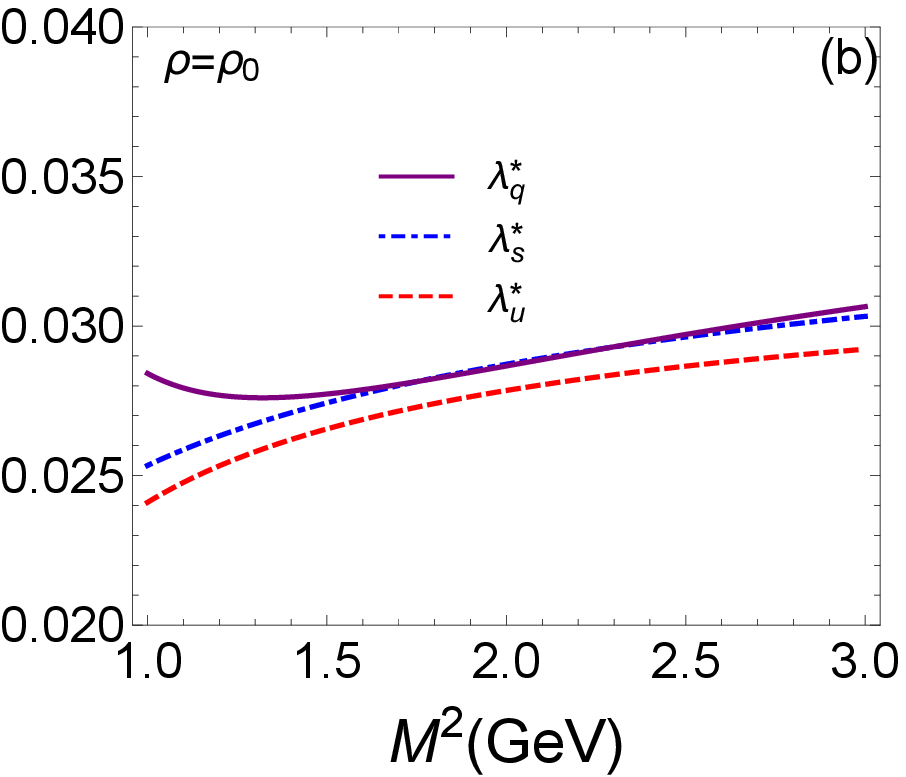} 
\caption{Borel curves for (a) the $\Delta$ isobar pole residues in vacuum and (b) the quasi-$\Delta$ isobar pole residues in symmetric medium at normal nuclear density. Units of vertical axis are $\textrm{GeV}^2$.}\label{fig2}
\end{figure}

We first examine the sum rules in vacuum. Considering the phenomenological structure~\eqref{phenvac1}, the ground state sum rules  can be written as
\begin{align}
\frac{1}{\pi } \int_{0}^{s_0}\, ds e^{-s /M^2}\textrm{Im} \left[  \Pi_{\mu\nu}(s) \right] &  \simeq  g_{\mu\nu}\left( \overline{\mathcal{W}}_M^{\textrm{subt.}}[\Pi_{\Delta,s}(q_0^2,\vert\vec{q}\vert)]  + \overline{\mathcal{W}}_M^{\textrm{subt.}}[\Pi_{\Delta,q}(q_0^2,\vert\vec{q}\vert)]\slash \hspace{-0.2cm} q \right)+ \cdots \nonumber\\
&\simeq  g_{\mu\nu}\left( -\lambda^2  m_{\Delta} e^{-m^2_{\Delta}/M^2}     -\lambda^2  e^{-m^2_{\Delta}/M^2} \slash \hspace{-0.2cm} q \right)+ \cdots,
\end{align}
where the irrelevant part, the structures proportional to $\gamma_\mu \gamma_\nu$, $p_\mu\gamma_{\nu}$ and so on, which contain the other spin projections, is omitted. Residue sum rules for each invariant can be independently expressed as
\begin{align}
  -\lambda_s^2 m_{\Delta} e^{-m^2_{\Delta}/M^2}& = \overline{\mathcal{W}}_M^{\textrm{subt.}}[\Pi_{\Delta,s}(q_0^2,\vert\vec{q}\vert)], \\
-\lambda_q^2  e^{-m^2_{\Delta}/M^2}&= \overline{\mathcal{W}}_M^{\textrm{subt.}}[\Pi_{\Delta,q}(q_0^2,\vert\vec{q}\vert)],
\end{align}
where the vacuum limit is taken for the weighted invariant.

If the $\Delta$ isobar couples  strongly to the interpolating current, the pole residues $\lambda_{s,q}$ should reflect a good Borel behavior and the two values extracted from their respective plateau should be similar. 
The two invariants have different dimensions and hence the spectral density in the Borel transformed dispersion relations have different weighting functions.  As we approximate the continuum part of the spectral density with a sharp step function at the threshold $s_0$,  it is natural to expect that a different threshold would correctly reflect the contributions from the physical continuum contribution in the respective sum rules.  Therefore, the continuum threshold is differently assigned as ${s^s_0}=(1.4~\textrm{GeV})^2$ for $\overline{\mathcal{W}}_M^{\textrm{subt.}}[\Pi_{\Delta,s}(q_0^2,\vert\vec{q}\vert)]$ and ${s^q_0}=(2.0~\textrm{GeV})^2$ for $\overline{\mathcal{W}}_M^{\textrm{subt.}}[\Pi_{\Delta,q}(q_0^2,\vert\vec{q}\vert)]$, respectively. 

The large difference in the continuum threshold for the two invariants can also be understood by considering the excited $\Delta$ states below 1.9 GeV with $I(J)=\frac{3}{2}(\frac{3}{2})$.  These are the two states $\Delta(1600)$ and $\Delta(1700)$ with positive and negative parity respectively.  
Due to the difference in the parities, the two states have the following phenomenological side:
\begin{align}
\Pi_{\mu\nu}(q) = \left( \lambda_+^2 \frac{\slash \hspace{-0.2cm} q + m_{+}}{q^2-m_+^2} +\lambda_-^2 \frac{\slash \hspace{-0.2cm} q - m_{-}}{q^2-m_-^2}\right)g_{\mu\nu} =
\slash \hspace{-0.2cm} q \bigg( \frac{\lambda_+^2 }{q^2-m_+^2} +\frac{\lambda_-^2 }{q^2-m_-^2}\bigg)g_{\mu\nu} +
  \left( \frac{\lambda_+^2  m_+}{q^2-m_+^2} -\frac{\lambda_-^2 m_-}{q^2-m_-^2}\right)g_{\mu\nu}
,\label{phenvac-pm}
\end{align}
where the subscript $\pm$ denotes the parity of the $\Delta$'s.  One notes that the two states contribute differently in the two polarization tensors and hence  will inevitably lead to  different effective thresholds.  Because of the finite truncation of the OPE, the scalar channel~\eqref{opes} does not reflect the cancellation tendency one can anticipate via Eq.~\eqref{phenvac-pm}. Thus limitation of the threshold at ${s^s_0}=(1.4~\textrm{GeV})^2$ is a plausible choice.  Furthermore, as the sum rule with different threshold leads to a stable sum rule for the $\Delta$ with experimental masses, the different threshold seems to guarantee the isolation of the ground state $\Delta$ particle in the pole structure.

\begin{figure}
\includegraphics[height=7.0cm]{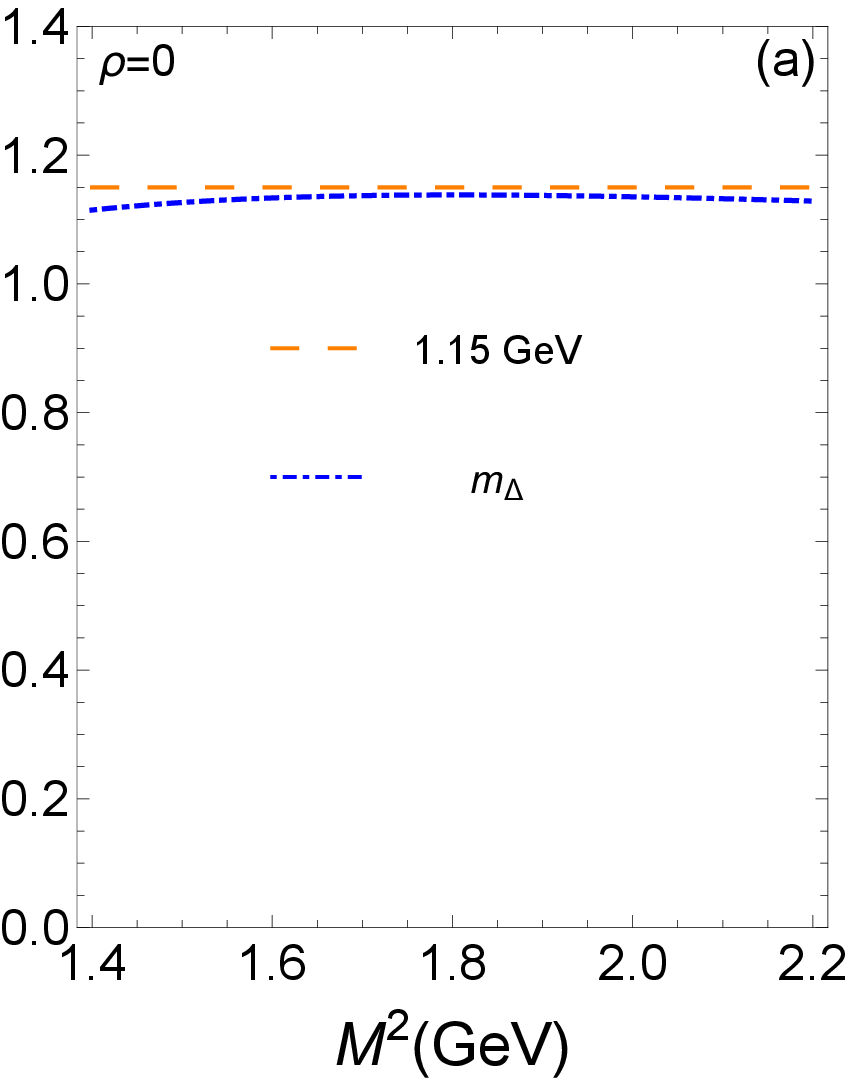} \quad
\includegraphics[height=7.0cm]{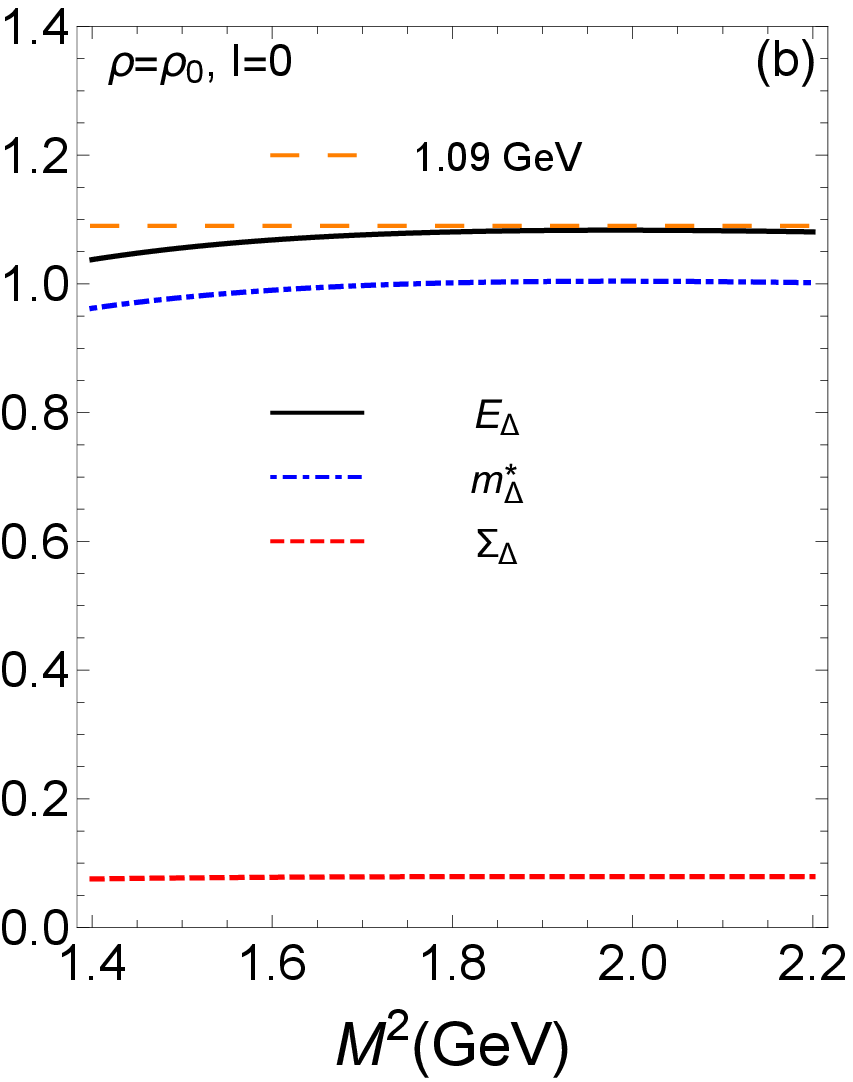} \quad
\includegraphics[height=7.0cm]{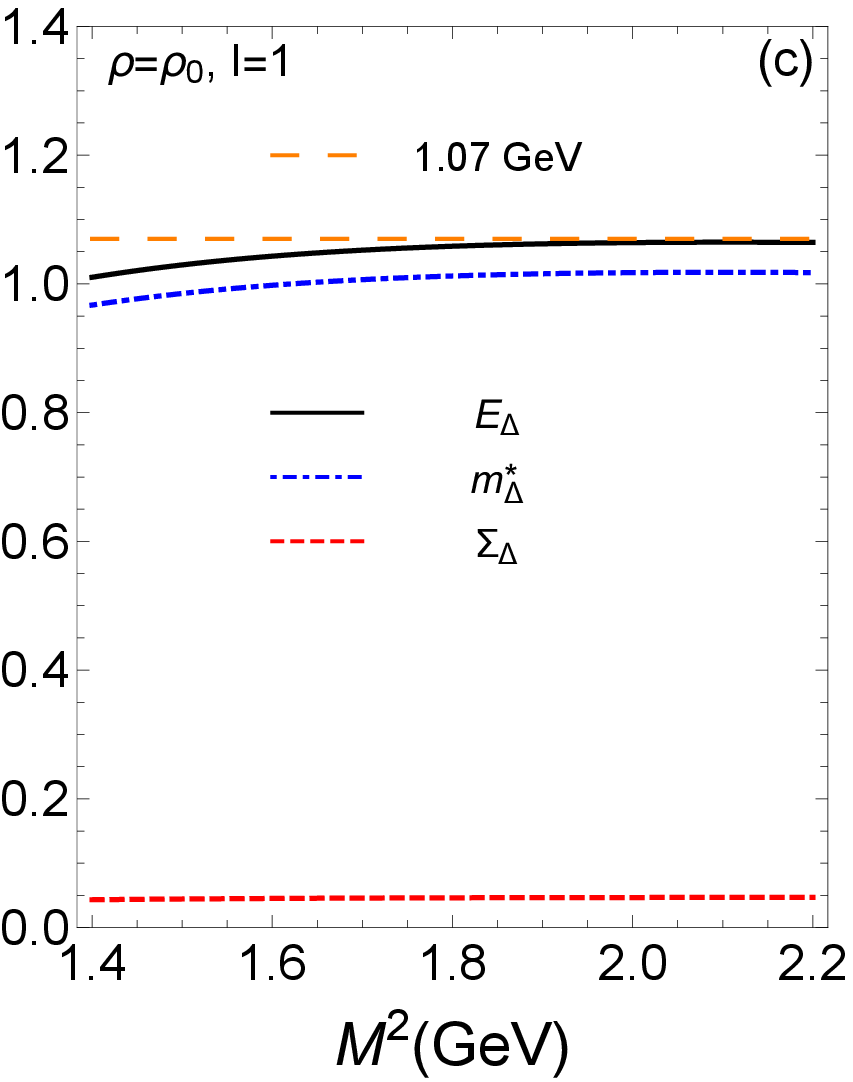} 
\caption{Borel curves for (a) the $\Delta$ isobar mass in vacuum, and (b) the quasi-$\Delta$ isobar self energies in symmetric matter, and (c) in neutron matter. (Orange) long-dashed lines represent reference values. Units of vertical axis are GeV.}
\label{fig3}
\end{figure}

As can be seen in Fig.~\ref{fig2}(a) both Borel curves show stable behavior in the proper Borel window with values close to each other within acceptable uncertainty. The corresponding mass curve is plotted in Fig.~\ref{fig3}(a) where one can read the $\Delta$ isobar mass as $m_{\Delta}\simeq 1.15~\textrm{GeV}$. 

We did a rough estimate of our errors by the following procedure: We varied the values of $s_0^s$ and $s_0^q$ by $(100~\textrm{MeV})^2$ above and below the value we used in the final analysis, as well as the value of $k_1$ by $0.1$ above and below $0.7$ and estimate the value of the error by taking the square root of the square of the variations of the vacuum mass of the $\Delta$ with these parameters. This procedure gives 22\% of error in our sum rule, most of which is coming from the variation of the scalar threshold. Errors of this magnitude are typical for QCD sum rule calculations and our results should be read with this in mind.

For the quasi-$\Delta$ isobar structure~\eqref{corrmed} in medium, one can write similar sum rules for the quasiparticle state as
 
\begin{align}
\frac{1}{2\pi } \int_{-\omega^{}_0}^{\omega^{}_0 } d\omega  & (\omega-\bar{E}_q)e^{-\omega^2/M^2}  \textrm{Im}\left[  \Pi_{\mu\nu}(\omega) \right] 
\nonumber\\
&\simeq  g_{\mu\nu}\left( \overline{\mathcal{W}}_M^{\textrm{subt.}}[\Pi_{\Delta,s}(q_0^2,\vert\vec{q}\vert)]  + \overline{\mathcal{W}}_M^{\textrm{subt.}}[\Pi_{\Delta,u}(q_0^2,\vert\vec{q}\vert)]\slash \hspace{-0.2cm} u  + \overline{\mathcal{W}}_M^{\textrm{subt.}}[\Pi_{\Delta,q}(q_0^2,\vert\vec{q}\vert)]\slash \hspace{-0.2cm} q \right)+ \cdots \nonumber\\
& \simeq g_{\mu\nu} \left( -{\lambda^{*}}^2 m^{*}_{\Delta} e^{-(m^{*}_{\Delta} +\Sigma_{v})^2/M^2} + {\lambda^{*}}^2  \Sigma_{v} e^{-(m^{*}_{\Delta}+\Sigma_{v})^2/M^2} \slash \hspace{-0.2cm} u -{\lambda^{*}}^2  e^{-(m^{*}_{\Delta}+\Sigma_{v})^2/M^2} \slash \hspace{-0.2cm} q \right)+ \cdots,\label{medsr}
\end{align}
where $m^{*}_{\Delta}=m_{\Delta} + \Sigma_{s} $. The analogous independent expression can be written as
\begin{align}
  -{\lambda^{*}_s}^2 m^{*}_{\Delta}  e^{-(m^{*}_{\Delta}+\Sigma_{v})^2/M^2} &=\overline{\mathcal{W}}_M^{\textrm{subt.}}[\Pi_{\Delta,s}(q_0^2,\vert\vec{q}\vert)] , \\
{\lambda^{*}_u}^2  \Sigma_{v}  e^{-(m^{*}_{\Delta}+\Sigma_{v})^2/M^2} &=  \overline{\mathcal{W}}_M^{\textrm{subt.}}[\Pi_{\Delta,u}(q_0^2,\vert\vec{q}\vert)] , \\
-{\lambda^{*}_q}^2    e^{-(m^{*}_{\Delta}+\Sigma_{v})^2/M^2} &=\overline{\mathcal{W}}_M^{\textrm{subt.}}[\Pi_{\Delta,q}(q_0^2,\vert\vec{q}\vert)].
\end{align}

As in the vacuum case, the  quasipole residues should be the same as they are defined from the same quasiparticle state.  Again, one can assign $\omega^{s}_0= \omega^{u}_0=1.5 ~\textrm{GeV}$ for $\overline{\mathcal{W}}_M^{\textrm{subt.}}[\Pi_{\Delta,s}(q_0^2,\vert\vec{q}\vert)]$ and $\overline{\mathcal{W}}_M^{\textrm{subt.}}[\Pi_{\Delta,u}(q_0^2,\vert\vec{q}\vert)]$, and $\omega^{q}_0= 2.0~\textrm{GeV}$ for $\overline{\mathcal{W}}_M^{\textrm{subt.}}[\Pi_{\Delta,q}(q_0^2,\vert\vec{q}\vert)]$. The residue sum rules are plotted in Fig.~\ref{fig2}(b), where the stability is less than in the vacuum reflecting the less distinct quasipole structure compared to the vacuum case.  Still, one finds moderate behavior of the residues for the scalar and medium vector invariant.  The quasiparticle self-energies in the isospin symmetric matter and the neutron matter are plotted in Figs.~\ref{fig3}(b) and  \ref{fig3}(c). In the symmetric condition, the strong attraction leads to $m^{*}_{\Delta}=m_{\Delta}+\Sigma_s\simeq 1.00~\textrm{GeV}$, which corresponds to a scalar attraction of about 150 MeV  and a  weak vector repulsion $\Sigma_{v} \simeq 75~\textrm{MeV}$, which lead to negative pole shift on the order of $\sim 75~\textrm{MeV}$ in comparison with the $\Delta$ isobar mass in vacuum. This reduction pattern is quite similar to the results of Refs.~\cite{Post:2003hu, Korpa:2003bc, Korpa:2008ut}.  In neutron matter,  the vector repulsion becomes even weaker  $\Sigma_{v} \simeq 0.045~\textrm{GeV}$ and the quasi-$\Delta^{++}$ state energy can be read as $\simeq 1.07~\textrm{GeV}$ corresponding to scalar attraction of 140 MeV, which is quite similar in magnitude to the quasi-neutron energy in  neutron matter~\cite{Jeong:2016qlk} . Finally, our result shows that the quasipole position, which corresponds to the  mass of the $\Delta^{++}$ in the medium, is reduced by 75 MeV (150 MeV scalar attraction and 75 MeV vector repulsion) in symmetric nuclear matter and by 95 MeV (140 MeV scalar attraction and 45 MeV vector repulsion) in neutron matter at nuclear matter density.

\subsection{Consideration of $\pi -N$ continuum state as ground state}

The interpolating field $\eta^{(2)}_{\mu}$ can couple to any state or continuum states  with the same quantum numbers as the $\Delta$ isobar. As the energy threshold of $\pi- N$ continuum state is expected to be lower than the mass of the $\Delta$ isobar, one may explicitly consider the  $\pi-N$ continuum like contribution. 
It should be noted that this contribution is present in the QCD sum rule approach, because one introduces an interpolating field that  couples not only to the $\Delta$ but also to the $\pi-N$ directly. 
Once this contribution is subtracted out, one can study the $\Delta$ property in medium within the sum rule approach, which can now be compared to any phenomenological calculation that includes the $\pi-N$ type of contribution  with short-range correlation effects such as introduced by the Migdal vertices~\cite{Nakano:2001ue,Lutz:2002qy}.   As mentioned before, our sum rule for the $\Delta$ indeed seems to give a result similar to the self-consistent phenomenological approach~\cite{Korpa:2008ut}.
Therefore, to identify and isolated the $\pi-N$ contribution that directly couples to the current, we introduce a  hybrid state to discriminate from the OPE continuum where the other excited states are expected to reside.

As briefly discussed in Sec.~\ref{sec2}, $\eta^{(2)}_{\mu}$ interpolates the isospin-$\frac{3}{2}$ state where the corresponding spin-$\frac{3}{2}$ ground state is expected. A phenomenological current for the hybrid state can be inferred from the following  vertex~\cite{Pascalutsa:1998pw, Almaliev:2002tg}:
\begin{align}
\mathcal{L}_{\textrm{int}}= g \epsilon^{\mu \nu \alpha \beta} \left( \partial_{\mu} \bar{\psi}_{\nu} \gamma_5 \gamma_{\alpha} \Psi_p \partial_{\beta} \pi + \bar{\Psi}_p \gamma_5 \gamma_{\alpha} \partial_{\mu} \psi_{\nu} \partial_{\beta} \pi  \right),
\end{align}
where $\Psi_p$ and $\pi$ represents proton and pion, respectively.  This vertex satisfies the transverse gauge condition $q^{\mu}\psi_{\mu} (q)  =0$ and the spin-$\frac{1}{2}$ part contained in RS field $\psi_{\mu}$ is excluded as shown in Ref.~\cite{Pascalutsa:1998pw}. Identifying the wave function of spin-$\frac{3}{2}$ state interpolated via $\eta^{(2)}_{\mu}$ to $\psi_\mu$, one can identify the $\pi-N$ current as
\begin{align}
J^{\mu}_{\pi N} (x) & =  \epsilon^{\mu\alpha  \nu \beta}  \gamma_5 \gamma_{\alpha} \partial_{\nu}  \Psi_p (x) \partial_{\beta} \pi(x).\label{pc}
\end{align}

Now one can calculate the hybrid ground state from the current~\eqref{pc} correlator:
 \begin{align}
\Pi_{\pi N}^{\mu \nu}(q) & \equiv i \int d^4 x e^{iqx} \langle\Psi_0\vert
\textrm{T}[J^{\mu}_{\pi N}(x) \bar{J}^{\nu}_{\pi N}(0)]\vert\Psi_0\rangle\nonumber\\
&= i   \slash \hspace{-0.2cm} q \left(  \mathcal{P}^{3/2}_{ } (q) - 2 \mathcal{P}^{1/2}_{11 } (q)\right)_{\mu \rho} \int \frac{d^4k}{(2\pi)^4} S^{(i)}_p(k) D_{\pi}(q-k) k^{\rho}  k^{\sigma}  \left(  \mathcal{P}^{3/2}_{ } (q) - 2 \mathcal{P}^{1/2}_{11 } (q)\right)_{\sigma \nu} \slash \hspace{-0.2cm} q,
\end{align}
where the hadron propagators are given as
\begin{align}
  S^{0}_p(k) &= \frac{\slash \hspace{-0.2cm} k+ m_p }{k^2 - m_p^2},\\
    S^{*}_p(k) &= \frac{\slash \hspace{-0.2cm} k -\Sigma^{p}_v \slash \hspace{-0.2cm} u+ m^{*}_p }{(k-\Sigma^{p}_v u)^2 - {m^{*}_p}^2},\\
    D_{\pi}(q-k) &=\frac{1}{(q-k)^2-m^2_{\pi}}.
\end{align}
The quasinucleon propagator $ S^{*}_p(k)$ is used with the self energies calculated in Refs.~\cite{Jeong:2012pa, Jeong:2016qlk}: $m^{*}_p= (0.6 + 0.05I )~\textrm{GeV}$, $\Sigma^{p}_v= (0.32 - 0.08I )~\textrm{GeV}$.  $ D_{\pi}(q-k)$ used in both the vacuum and the in-medium calculation as the in-medium correction is expected to be small in the chiral limit ($m_{\pi} \rightarrow 0$). The weighted invariant can be summarized as
\begin{align}
\overline{\mathcal{W}}_M^{\textrm{subt.}}[ \Pi_{\pi N}^{\mu \nu}(q)] &\Rightarrow  \left(  \mathcal{P}^{3/2}_{ } (q) + 4 \mathcal{P}^{1/2}_{11 } (q)\right)^{\mu \nu} \frac{1}{16 \pi}\int_{m_p^{*}+\Sigma^p_v}^{\omega^{ }_0} d\omega e^{-\omega^2/M^2}\left(-\textrm{Im}\Pi_{q}(\omega) \slash \hspace{-0.2cm} q -\textrm{Im}\Pi_{u}(\omega) \slash \hspace{-0.2cm} u - \textrm{Im}\Pi_{s}(\omega) \right)+\cdots \nonumber\\
& \equiv \left(  \mathcal{P}^{3/2}_{ } (q) + 4 \mathcal{P}^{1/2}_{11 } (q)\right)^{\mu \nu} \left( -\rho_q(M^2) \slash \hspace{-0.2cm} q - \rho_u (M^2) \slash \hspace{-0.2cm} u - \rho_s(M^2)  \right)+\cdots,
\end{align}
where only $\mathcal{P}^{3/2}_{ } (q) + 4 \mathcal{P}^{1/2}_{11 } (q)$ proportional invariants are kept and the  $\textrm{Im}\Pi_{i}(\omega) $ are calculated  to be
\begin{align}
\textrm{Im}\Pi_{q}(\omega)& = \frac{(\omega^2)^2}{24} +  \frac{({m_p^{*}}^2)^3}{12 \omega^2}-  \frac{{m_p^{*}}^2  \omega^2}{12}    -    \frac{({m_p^{*}}^2)^4}{24 (\omega^2)^2}\nonumber\\
& \qquad - (\Sigma^p_v)^2 \left(  \frac{\omega^2}{24} +   \frac{({m_p^{*}}^2)^3}{ 12(\omega^2)^2}-  \frac{{m_p^{*}}^2}{12}    -   \frac{({m_p^{*}}^2)^4}{24 (\omega^2)^3}\right),\\
\textrm{Im}\Pi_{u}(\omega)& =\Sigma^p_v \left( \omega + \Sigma^p_v \right)^2 \left(   \frac{\omega^2}{24}  +    \frac{({m_p^{*}}^2)^3}{12(\omega^2)^2}-  \frac{ {m_p^{*}}^2  }{12}  -    \frac{({m_p^{*}}^2)^4}{ 24(\omega^2)^3}\right),\\
\textrm{Im}\Pi_{s}(\omega)& = m_p^{*} \left( \omega + \Sigma^p_v \right)^2 \left(   \frac{\omega^2}{12}  +    \frac{({m_p^{*}}^2)^2}{4\omega^2 }-  \frac{ {m_p^{*}}^2  }{4}  -    \frac{({m_p^{*}}^2)^3}{ 12(\omega^2)^2}\right).
\end{align}
The weighted invariants reduce to the vacuum structure in the  $u_{\mu} \rightarrow 0$ and $\rho\rightarrow 0 $ limits.
By defining the $\eta^{(2)}_{\mu}$-hybrid coupling strength as $c_{\pi N}$, one can rewrite the sum rules for the ground state~\eqref{medsr} as 
\begin{align}
\frac{1}{2\pi } \int_{-\omega^{*}_0}^{\omega^{*}_0} d\omega  & (\omega-\bar{E}_q)e^{-\omega^2/M^2}  \textrm{Im}\left[  \Pi_{\mu\nu}(\omega) \right] 
\nonumber\\
& \simeq g_{\mu\nu} \bigg( -{\lambda^{*}}^2 m^{*}_{\Delta} e^{-(m^{*}_{\Delta} +\Sigma_{v})/M^2} + {\lambda^{*}}^2  \Sigma_{v} e^{-(m^{*}_{\Delta}+\Sigma_{v})/M^2} \slash \hspace{-0.2cm} u -{\lambda^{*}}^2  e^{-(m^{*}_{\Delta}+\Sigma_{v})/M^2} \slash \hspace{-0.2cm} q \nonumber\\
&\qquad\qquad   - c_{\pi N}^2 \rho_s(M^2) - c_{\pi N}^2 \rho_u (M^2) \slash \hspace{-0.2cm} u- c_{\pi N}^2  \rho_q(M^2) \slash \hspace{-0.2cm} q\bigg),\label{medsr2}
\end{align}
which leads to following sum rules:
\begin{align}
 -{\lambda^{*}_s}^2 m^{*}_{\Delta}  e^{-(m^{*}_{\Delta}+\Sigma_{v})/M^2} &=\overline{\mathcal{W}}_M^{\textrm{subt.}}[\Pi_{\Delta,s}(q_0^2,\vert\vec{q}\vert)]+ c_{\pi N}^2 \rho_s(M^2)  , \\
{\lambda^{*}_u}^2  \Sigma_{v}  e^{-(m^{*}_{\Delta}+\Sigma_{v})/M^2} &=  \overline{\mathcal{W}}_M^{\textrm{subt.}}[\Pi_{\Delta,u}(q_0^2,\vert\vec{q}\vert)]+ c_{\pi N}^2 \rho_u(M^2)  , \\
-{\lambda^{*}_q}^2    e^{-(m^{*}_{\Delta}+\Sigma_{v})/M^2} &=\overline{\mathcal{W}}_M^{\textrm{subt.}}[\Pi_{\Delta,q}(q_0^2,\vert\vec{q}\vert)]+ c_{\pi N}^2 \rho_q(M^2) .
\end{align}

\begin{figure}
\includegraphics[width=6.3cm]{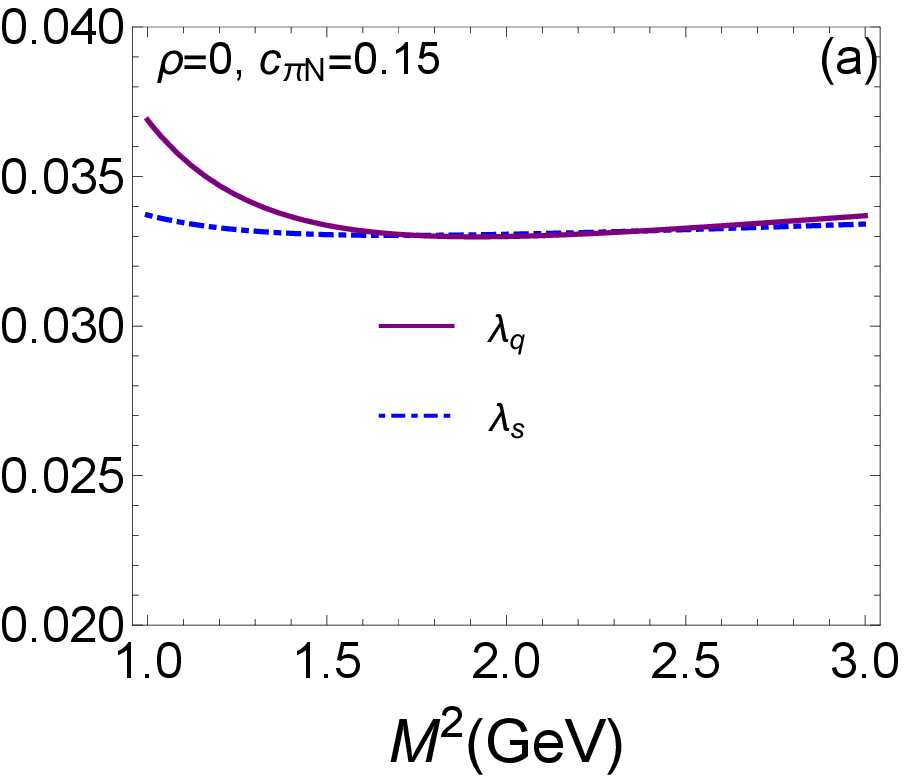} \quad
\includegraphics[width=6.3cm]{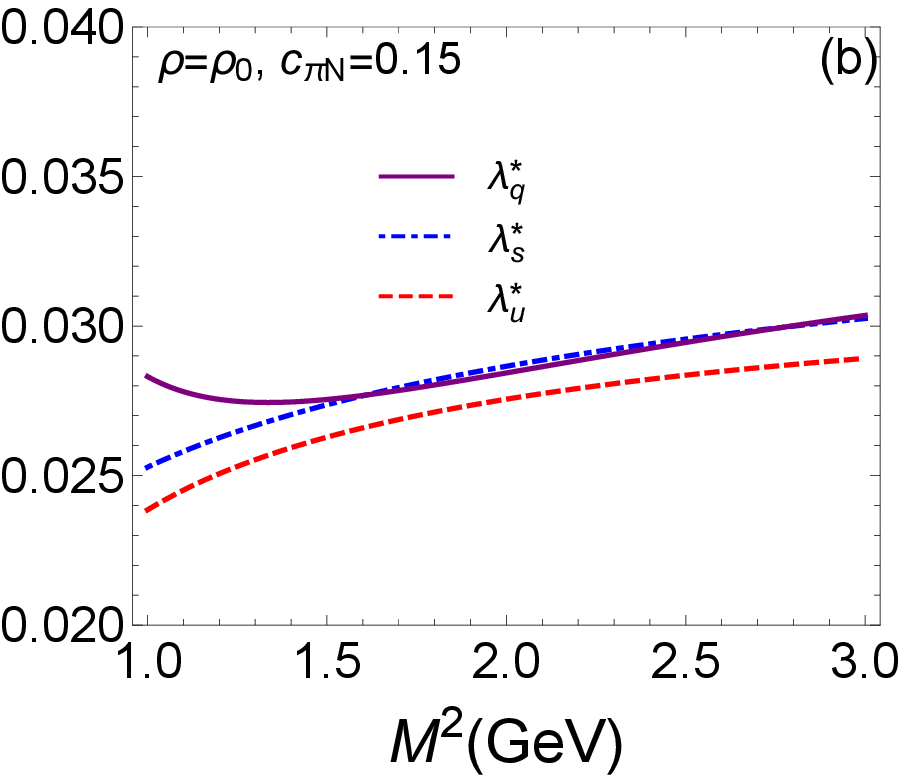} 
\caption{$\pi-N$ continuum subtracted Borel curves for (a) the $\Delta$ isobar pole residues in vacuum and (b) the quasi-$\Delta$ isobar pole residues in medium. Units of vertical axis is $\textrm{GeV}^2$ and $c_{\pi N}=0.15$.}\label{fig4}
\end{figure}

\begin{figure}
\includegraphics[height=7.0cm]{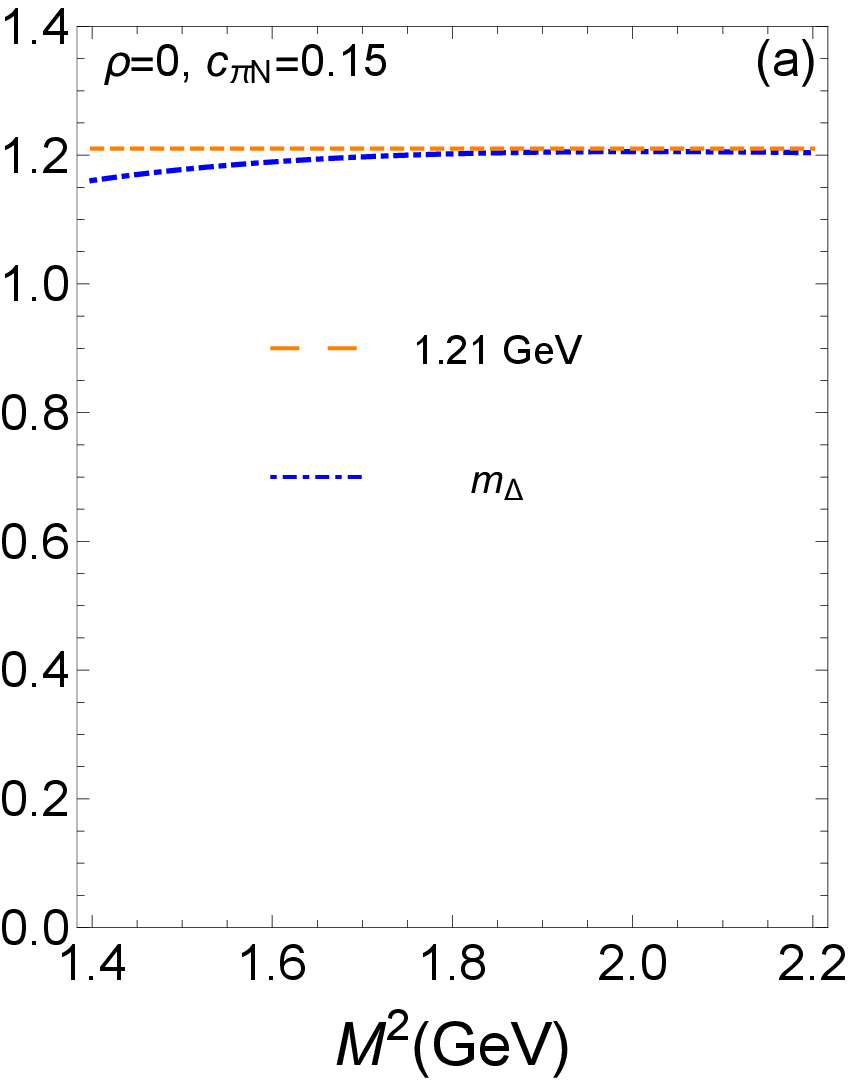} \quad
\includegraphics[height=7.0cm]{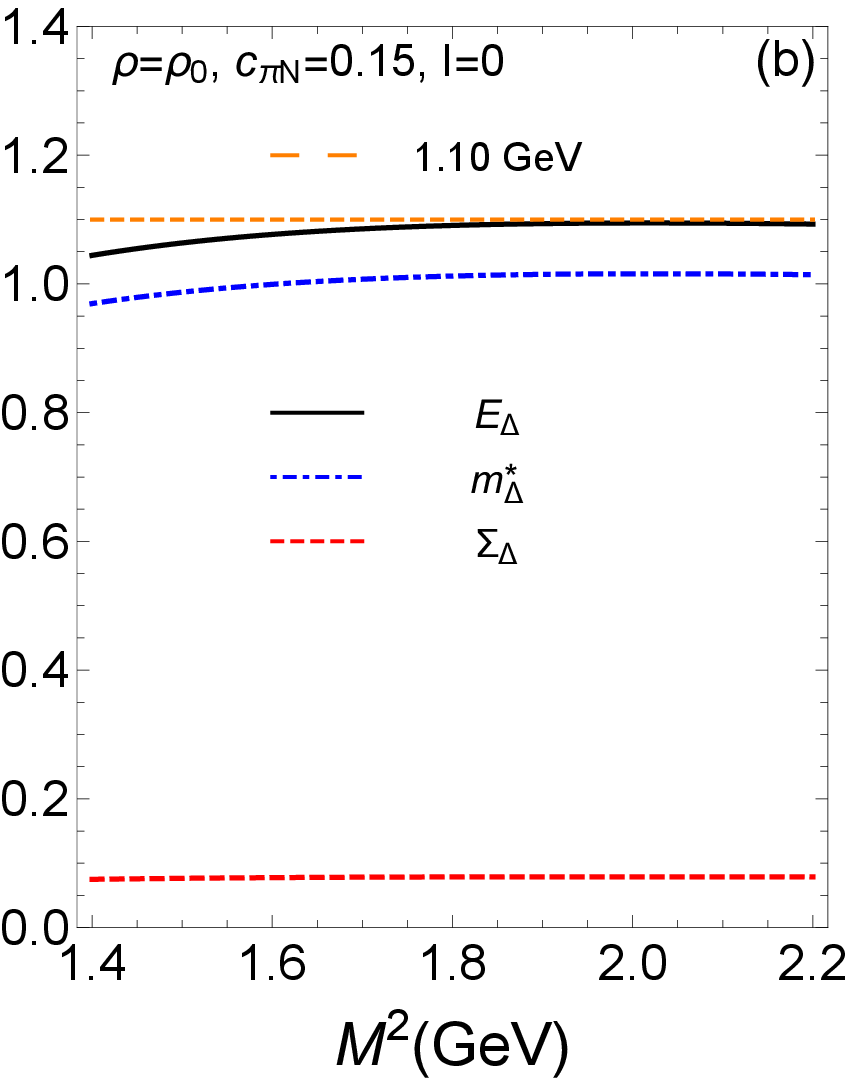} \quad
\includegraphics[height=7.0cm]{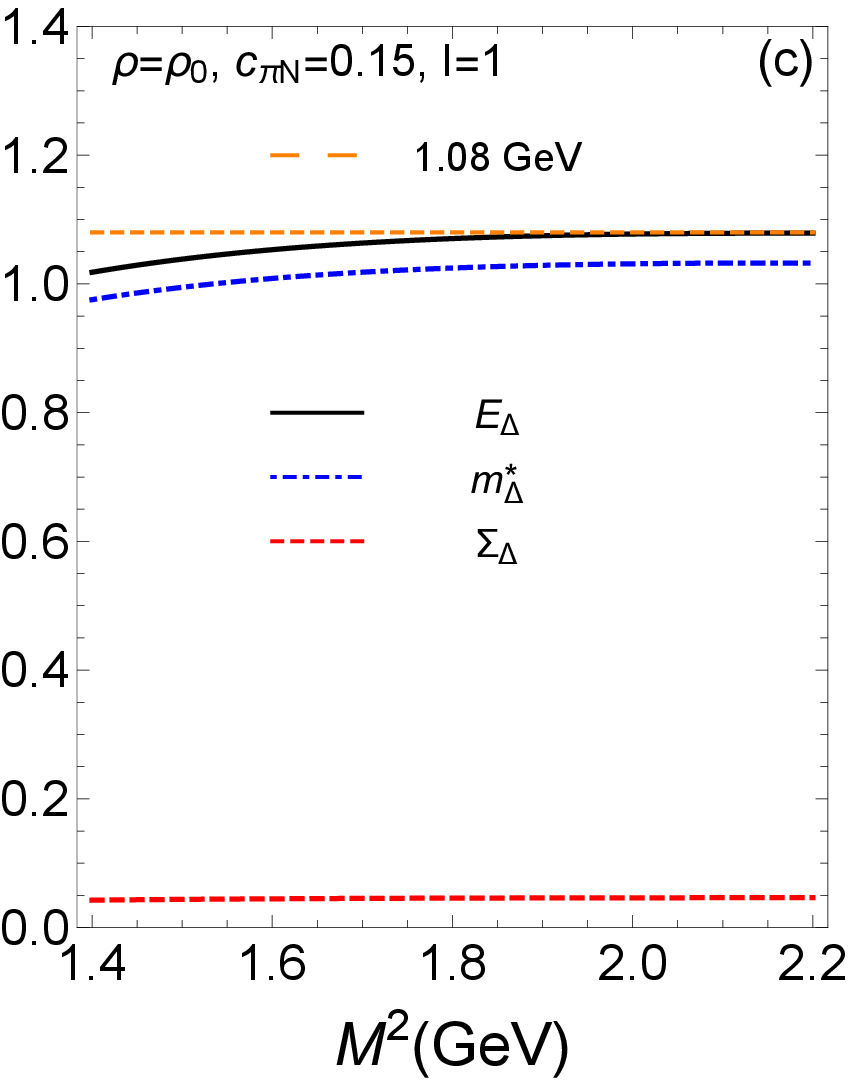} 
\caption{$\pi-N$ continuum subtracted Borel curves for (a) the $\Delta$ isobar mass in vacuum, (b) the quasi-$\Delta$ isobar self energies in symmetric matter, and (c) in neutron matter. Orange long-dashed lines represent reference values.  Units of vertical axis is  $\textrm{GeV}$ and and $c_{\pi N}=0.15$.}\label{fig5}
\end{figure}

As plotted in Fig.~\ref{fig4}(a), the Borel curves for the pole residues do not change  drastically  but provide a better  behavior in the vacuum sum rules. Although the hybrid contribution is minimal in the range of $0.0 \leq c_{\pi N} \leq 0.2$, the subtraction of this hybrid like structure makes the pole contribution even more clear with $m_{\Delta}=1.21~\textrm{GeV}$ [Fig.~\ref{fig5}(a)]. However, the same  subtraction scheme does not change the in-medium sum rules very much.
However, it should be pointed out that if the softening of the pion spectrum is taken into account, the situation might change.  This is so because although the direct $\pi-N$ continuum is a contribution appearing in the sum rule approach, it is nevertheless correlated to the continuum appearing in the $\Delta$ self-energy as the total spectral density in the correlation function is identified to the changes in the operator product expansion.
The quasipole residues plotted in Fig.~\ref{fig4}(b) and the self-energies plotted in the curves in Figs.~\ref{fig5}(b) and \ref{fig5}(c) are almost unchanged from the plots in Figs.~\ref{fig2}(b), ~\ref{fig3}(b), and \ref{fig3}(c), respectively. The negative pole shift becomes of order  100 MeV (200 MeV scalar attraction and 75 MeV vector repulsion in the symmetric matter), which agrees with the experimental observation~\cite{Pelte:1999ir}.

The other apparent feature is that the scalar self-energy is almost isospin independent [Figs.~\ref{fig3}(c) and  \ref{fig5}(c)]. This tendency agrees with the phenomenology discussed in Ref.~\cite{Cai:2015hya} where the quasibaryon vector self-energy has been arranged as 
\begin{align}
\Sigma_{v} &\equiv g_{\omega \Delta}\bar{\omega}_0 + \tau^{3}_{i}  g_{\rho \Delta}\bar{\rho}^{(3)}_0,\\
\Sigma^N_{v} &\equiv g_{\omega N}\bar{\omega}_0 + \tau^{3}_{p/n}  g_{\rho N}\bar{\rho}^{(3)}_0,
\end{align}
where $g_{i}$ represents the meson exchange coupling and $ \tau^{3}_{p/n}=\pm1$. Following this notation, the vector self-energy in this work can be expressed as
\begin{align}
\Sigma_{v}(\rho, I) &= - \frac{ \overline{\mathcal{W}}_M^{\textrm{subt.}}[\Pi_{\Delta,u}(q_0^2,\vert\vec{q}\vert)] }{\overline{\mathcal{W}}_M^{\textrm{subt.}}[\Pi_{\Delta,q}(q_0^2,\vert\vec{q}\vert)]} \simeq \Sigma_{v}(\rho_0, 0) + \left( \Sigma_{v}(\rho_0, 1)-\Sigma_{v}(\rho_0, 0) \right)I \simeq (0.075 -0.010 \tau^{3}_{\Delta} I )~\textrm{GeV},
\end{align}
where $\tau^{3}_{i}$ is defined as $\tau^{3}_{\Delta^{++}}=3$, $\tau^{3}_{\Delta^{+}}=1$, $\tau^{3}_{\Delta^{0}}=-1$, and $\tau^{3}_{\Delta^{-}}=-3$. If one considers the proton self-energies calculated in  Refs.~\cite{Jeong:2012pa, Jeong:2016qlk}, where $\Sigma^{p}_v= (0.32 - 0.08I )~\textrm{GeV}$, the ratio $x_\rho \equiv  g_{\rho \Delta}/g_{\rho N}$ becomes very small ($x_\rho \simeq 0.13$), which leads to the early appearance of $\Delta$ isobar in the dense medium~\cite{Cai:2015hya}. 

\section{Weak vector repulsion in a constituent quark model}\label{sec4}

 Our result shows that the vector repulsion for the $\Delta$ in medium is weaker than that for the nucleon.  This result can be understood by making use of the Pauli principle and constituent quark model. In the quark cluster model, the short-range vector repulsion can be shown to arise from combining Pauli principle and the quark two-body interactions between quarks~\cite{Oka:1980ax, Oka:1981ri}.  The repulsion can be estimated by comparing the static energy of dibaryon configuration to that of the two separated baryon states. 

If we represent the dibaryon state using coefficients of fractional parentage (cfp)~\cite{Harvey:1988nk}, then we can estimate how much baryon-baryon state is included in dibaryon state with specific quantum number. If we only consider an $s$-wave dibaryon state so that the orbital wave function is totally symmetric,  the corresponding dibaryon states are as follows. 

For $(I,S)$=(1,0) or (0,1),
\begin{align}
  \Psi=\frac{1}{3}NN+\frac{2}{3\sqrt{5}}\Delta\Delta+\frac{2}{\sqrt{5}}CC,  \label{NN}
\end{align}
where $CC$ means hidden color states that cannot be represented in terms of free baryons.

For $(I,S)$=(2,1) or (1,2),
\begin{align}
  \Psi=\frac{1}{3}\Delta\Delta + \frac{2}{3\sqrt{5}}N\Delta + \frac{2}{\sqrt{5}}CC, \label{DeltaN}
\end{align}
where the relative orbital state for $N$ and $\Delta$ is symmetric. 
These compositions determine the fractional weight of  repulsion in the dibaryon configuration that contributes to the respective repulsion in the two-baryon interaction at short distance.  
It can be shown that the difference in the interaction energy between the dibaryon configurations and two separated baryons are dominated by the color-spin interaction, while all other $s$-wave interaction strength cancel~\cite{Park:2018ukx, Park:2015nha}.  Assuming that the dibaryon and the two-baryon occupy the same spatial configurations, the color spin interaction strength can be shown to be proportional to the following matrix element:
\begin{eqnarray}
V=-\sum^n_{i<j}\lambda^c_i \lambda^c_j   \sigma_i \cdot \sigma_j \label{hyperfine}
\end{eqnarray}

In Table~\ref{table-1}, we show the expectation value of Eq.~(\ref{hyperfine}) for dibaryon  in the first row with the difference between the dibaryon and the two lowest energy baryons in the second row. As we can see in Table~\ref{table-1}, all the dibaryon states show repulsive interaction except $(I=0,S=3)$, which corresponds to the two-$\Delta$ decay channel.  Hence, there is always repulsion between two nucleons or between the $\Delta$-nucleon states. 

We can compare the strength of the repulsion   between  $NN$ and $N\Delta$ by calculating the fractional contribution of the dibaryon configuration to their respective channels.  

\begin{table}
\addtolength{\tabcolsep}{+15pt}
  \begin{tabular}{ccccccc}
    \hline
    \hline
    ($I,S$) & (3,0) & (2,1) & (1,2) & (1,0) & (0,3) & (0,1) \\
    \hline
    $V_d$ & 48 & $\frac{80}{3}$ & 16 & 8 & 16 & $\frac{8}{3}$ \\
    \hline
    $\Delta V$ & 32 & $\frac{80}{3}$ & 16 & 24 & 0 & $\frac{56}{3}$ \\
    \hline
    \hline
  \end{tabular}
  \caption{$V_d$ is the expectation value of $-\sum^6_{i<j}\lambda^c_i \lambda^c_j  \sigma_i \cdot \sigma_j $. $\Delta V$ is $V_d-(V_{b1}+V_{b2})$, in which $V_{b1}$ and $V_{b2}$ are those baryons to which the dibaryon can decay.}
\label{table-1}
\end{table}

\begin{enumerate}
  \item{Spin and isospin-averaged $N$-$N$ interaction:}\\
Among the possible states in Table~\ref{table-1}, only two $(I,S)$ states, which are $(0,1)$ and $(1,0)$, contain $NN$ states.  Making use of the $NN$ component in the respective dibaryon configuration given in Eq.~(\ref{NN}), we find the relative repulsion to be as follows:
\begin{align}
\frac{1}{2} \bigg( H_{0,1}^{NN}+H_{1,0}^{NN} \bigg)=\frac{1}{2} \left(\frac{1}{3}\right)^2 \frac{56}{3} + \frac{1}{2} \left(\frac{1}{3}\right)^2 24 = \frac{64}{27} \simeq 2.37.
\end{align}

  \item{Spin- and isospin-averaged $N$-$\Delta$ interaction:}\\
Similar to $NN$ case, only two $(I,S)$ states, which are $(1,2)$ and $(2,1)$, contain $N\Delta$ state. Making use of the $N-\Delta$ component in the respective dibaryon configuration given in Eq.~(\ref{DeltaN}), we find the relative repulsion to be as follows:
\begin{align}
\frac{1}{2} \bigg(  H_{1,2}^{N\Delta}+H_{2,1}^{N\Delta} \bigg)=\frac{1}{2} \left(\frac{2}{3\sqrt{5}}\right)^2 16 + \frac{1}{2} \left(\frac{2}{3\sqrt{5}}\right)^2 \frac{80}{3} =\frac{256}{135} \simeq 1.9.
\end{align}

\end{enumerate}

Therefore, we can conclude that the repulsion in $N\Delta$ is 20\% smaller than that in $NN$. This trend of a  weak vector repulsion from the $\Delta$ isobar is consistent with the above QCD sum rule results.

\section{Discussion and Conclusions}\label{sec5}

In this work, we calculated the quasi-$\Delta$ isobar energy in the isospin asymmetric matter. We allowed for different continuum  thresholds for the invariants with different dimensions and obtained an stable Borel curve for the $\Delta$ isobar mass in the vacuum. 
The quasi-$\Delta^{++}$ self energies in the medium are also  obtained within the stable  Borel curves.  We find that the vector self energy in the medium, which can be understood as the repulsive vector potential in the mean field approximation, is very weak in comparison to the nucleon case~\cite{Drukarev:1988kd, Cohen:1991js, Cohen:1996sb, Cohen:1994wm, Jeong:2012pa, Jeong:2016qlk}. As an order of magnitude estimate, the attraction in the scalar channel is 200 MeV and the repulsion in the vector channel is less than 100 MeV.

As the interpolating field $\eta^{(2)}_{\mu}$ can couple to the $\pi-N$ continuum state with energy threshold lower than the $\Delta^{++}$ mass, we also explicitly considered the phenomenological structure coming from this continuum. For the vacuum sum rules, the subtraction of $\pi -N$ continuum makes the pole signal more prominent, whereas in the medium case, it does not make any significant changes with respect to the quasi-$\Delta^{++}$ sum rules without the $\pi - N$ continuum. 
The situation might change if the 
softening of the pion spectrum is taken into account.  This is so because although the direct $\pi-N$ continuum is a contribution appearing in the sum rule approach, it is nevertheless correlated to the broadening appearing in the $\Delta$ self-energy as the total spectral density in the correlation function is identified to the changes in the operator product expansion.  
The broadening of the $\Delta$ can be found in studies that consistently take into account the quasiparticle and the quasiparticle loop structure~\cite{Post:2003hu, Korpa:2003bc} within the mean field self-energies. 
In particular, it would be extremely useful to find an improved interpolating current with which one can identify the part of the OPE in the $\Delta$ correlation function that corresponds to explicit structures of a quasiparticle  such as the Migdal contact interaction vertex~\cite{Nakano:2001ue, Lutz:2002qy, Korpa:2008ut}. We will leave the subject for a future discussion and research for the QCD sum rule study in the spin-$\frac{3}{2}$ state.
 
In the neutron matter, the quasiparticle energy is found to be similar to that in the isospin symmetric condition. This is because the scalar self-energy does not depend strongly on the isospin asymmetry and the vector self-energy is weak. If this attractive tendency is still valid in the dense regime, the stiff symmetry energy shown in Ref.~\cite{Jeong:2012pa} would allow $nn \leftrightarrow p \Delta^{-}$ in the equilibrium. This strong attraction would lead to a soft equation of state even if there is no hyperon condensation in the matter as discussed in Ref.~\cite{Cai:2015hya}. 

Although the attractive tendency of the quasi-particle energy agrees with the experimental observations~\cite{Barrette:1994kq,Hjort:1997wk,Pelte:1999ir,Eskef:2001qg,Bi:2005ca}, a  strong scalar attraction with weak vector repulsion seems problematic in relation to the recently established maximum neutron star mass which is close to $2  {\rm M_{\odot}}$~\cite{Demorest:2010bx, Antoniadis:2013pzd}. In the mean field approximation, the potential can be understood as being the average interaction between the quasi-$\Delta$ isobar and the  surrounding nucleons, namely the averaged two-particle interaction.  A strong three-body repulsion would provide a  possible resolution to the problem~\cite{Park:2018ukx}.  
A three-body repulsion between the $\Delta$ and two nucleons at short distance will only become important when the density becomes large.  Then the linear density extrapolation of strong scalar attraction will be saturated by the higher density repulsive force which will prevent the $\Delta$ condensation from occuring.  

\acknowledgments
This material is based upon work supported by
the São Paulo Research Foundation (FAPESP) under
Grants No. 2017/15346-0 and No. 2016/02717-8 (J.M.L. and R.D.M)
and Korea National Research Foundation
Grants. No. 
NRF-2016R1D1A1B03930089 (S.H.L.), No. NRF-2018R1D1A1B07043234 (A.P.), and No. NRF-2017R1D1A1B03033685 (K.S.J.). A part of the calculation of the OPE was checked using the package ``FeynCalc 9.0'' ~\cite{Mertig:1990an, Shtabovenko:2016sxi}.

\appendix

\section{Interpolating currents for the $\Delta$ isobar}\label{apena}
\subsection{Possible interpolating currents and relations between them}
For the purpose of describing the $\Delta$ within QCD sum rules, we initially considered the following set of currents in the $\left( \frac{1}{2}, 1 \right) \oplus \left(1,  \frac{1}{2} \right) $ representation:
\begin{align} 
	\eta^{(1)}_{\mu(1,2)3} &= (q_1^T C \sigma^{\alpha\beta} q_2)  \sigma_{\alpha\beta}  \gamma_{\mu} q_3 ,\\
	\eta^{(2)}_{\mu(1,2)3} &=  (q_1^T C \gamma_{\mu} q_2) q_3 ,\\
	\eta^{(3)}_{\mu(1,2)3} &=  (q_1^T C \gamma^{\nu} q_2)\sigma_{\mu\nu} q_3 ,\\
	\eta^{(4)}_{\mu(1,2)3} &=  (q_1^T C \sigma_{\mu\nu} q_2)\gamma^{\nu} q_3,
\end{align}
where $q_1$, $q_2$ and $q_3$ are one of the light quark fields, according to the hadron we want to describe. If $q_1=q_2=q_3=u$, that is, when all quark fields are the ones for the up quark, the sum rule calculated with this current should describe the $\Delta^{++}$ resonance, and respectively when they are all down quark fields, this should describe the $\Delta^-$. In this case, with all quarks being the same, it is possible to show through Fierz rearrangement that
\begin{equation}
\eta^{(1)}_{\mu}=4\eta^{(2)}_{\mu}=4i\eta^{(3)}_{\mu}=-4i\eta^{(4)}_{\mu},
\label{eq:reluuu}
\end{equation}
so the current choice is unique in this representation up to an overall numerical factor.
If one of the quark fields in the current corresponds to a different flavor from the other two, the currents should couple to the $\Delta^+$ (two up quarks and one down) and the $\Delta^0$ (two down quarks and one up). In this case, however, taking the example for the $\Delta^+$ current, there are two possible configurations, $\eta_{\mu(u,u)d}$ and $\eta_{\mu(u,d)u}$. Again, through Fierz rearrangement, one proves
\begin{align} 
	\eta^{(1)}_{\mu(u,d)u} &= 3\eta^{(2)}_{\mu(u,u)d}+i\eta^{(3)}_{\mu(u,u)d} ,\\
	\eta^{(2)}_{\mu(u,d)u} &= \frac{1}{8} \eta^{(1)}_{\mu(u,u)d} +\frac{1}{4} \eta^{(2)}_{\mu(u,u)d}+\frac{i}{4} \eta^{(3)}_{\mu(u,u)d} ,\\
	\eta^{(3)}_{\mu(u,d)u} &= -\frac{i}{8} \eta^{(1)}_{\mu(u,u)d} -\frac{3i}{4} \eta^{(2)}_{\mu(u,u)d}-\frac{1}{4} \eta^{(3)}_{\mu(u,u)d} ,\\
	\eta^{(4)}_{\mu(u,d)u} &= \frac{i}{8} \eta^{(1)}_{\mu(u,u)d} +\frac{3i}{4} \eta^{(2)}_{\mu(u,u)d}-\frac{1}{4} \eta^{(3)}_{\mu(u,u)d}-\frac{1}{2}\eta^{(4)}_{\mu(u,u)d}.
\end{align}

From these relations one can show that
\begin{equation}
\eta^{(1)}_{\mu(u,u)d}+2\eta^{(1)}_{\mu(u,d)u}=4\left(\eta^{(2)}_{\mu(u,u)d}+2\eta^{(2)}_{\mu(u,d)u}\right)=4i\left(\eta^{(3)}_{\mu(u,u)d}+2\eta^{(3)}_{\mu(u,d)u}\right)=-4i\left(\eta^{(4)}_{\mu(u,u)d}+2\eta^{(4)}_{\mu(u,d)u}\right),
\end{equation}
which is the generalization of the relations between the currents when all flavors are the same, Eq. \ref{eq:reluuu}. As explained in Appendix \ref{appendrenorm}, the current $\eta^{(1)}_\mu$ is renormalization covariant, irrespective of the quark flavors. Hence we find that the combination $(2\eta^{(i)}_{\mu(u,d)u}+\eta^{(i)}_{\mu(u,u)d})$, for $i=1$ to $4$, is renormalization covariant, since this is true for $i=1$ and they are all proportional to each other.

If one considers explicit light quark flavors, the  $  \left(\frac{1}{2} ,1 \right) \oplus \left(1,  \frac{1}{2} \right) $ representation contains the isospin-$\frac{1}{2}$ configuration which corresponds to the nucleon state, so simply using the current $\eta^{(i)}_{\mu(u,d)u}$ or $\eta^{(i)}_{\mu(u,u)d}$ is not enough to ensure that we are describing the $\Delta^+$ state, because there can be mixing with the proton spectral density.

We expected that the sum rule done with the current $(2\eta_{\mu(u,d)u}+\eta_{\mu(u,u)d})$ would be describing the $\Delta^+$ only, since this configuration seems to be the natural generalization of $\eta_{\mu(u,u)u}$. However, the requirement that in the vacuum and symmetric nuclear matter that the sum rule obtained using this current and the one using $\eta_{\mu(u,u)u}$ be the same, from the isospin symmetry of these systems, was not fulfilled. The issued is that although the OPE obtained with the $(2\eta_{\mu(u,d)u}+\eta_{\mu(u,u)d})$ current is proportional to the one obtained with the $\eta^{(i)}_{\mu(u,u)u}$ up to dimension 4 operators, for the dimension 6, the 4-quark operators, the situation seems to be more subtle. We intend to address this puzzle in a subsequent work and in this article we focused on the case for $\Delta^{++}$ and $\Delta^-$ only.

\subsection{The renormalization of the interpolating currents}\label{appendrenorm}
\begin{figure}\center
\includegraphics[height=3.7cm]{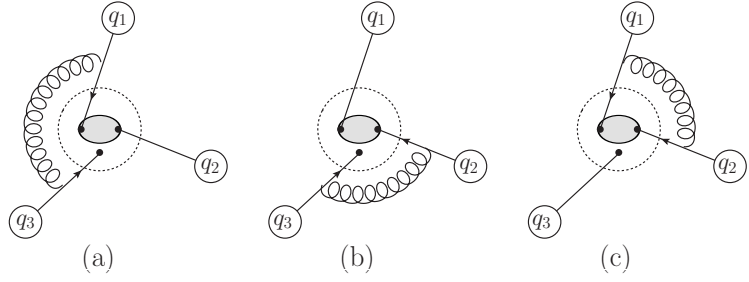} 
\caption{Anomalous correction of the interpolating current in one-loop order. Shaded ellipse in the dotted circle represents the diquark in the interpolating current.}\label{fig6}
\end{figure}

As introduced in Sec.~\ref{sec2}, the interpolating current for $\Delta^{++}$ is uniquely determined in the $\left( \frac{1}{2}, 1 \right) \oplus \left(1,  \frac{1}{2} \right) $ representation. However, for the $\Delta^{+}$ and $\Delta^{0}$, the choice is non-trivial as the local operator itself contains isospin-$\frac{1}{2}$ configuration and, also, the anomalous running can be different by the choice of local composite operator.

Consider the first type of interpolating field with explicit light quark flavors. 
\begin{align}  
\eta^{(1)}_{\mu(12)3}&\equiv   (q_1^T C \sigma^{\alpha\beta} q_2)  \sigma_{\alpha\beta}  \gamma_{\mu}q_3, 
\end{align}
where the numbers in parentheses represent the light quark flavors in the diquark. Then, the one-loop corrections can be calculated as presented in Fig.~\ref{fig6}:
\begin{align}  
\delta_{(a)} \left[ \eta^{(1)}_{\mu(12)3} \right] + \delta_{(b)} \left[ \eta^{(1)}_{\mu(12)3} \right] =  \frac{\alpha_s}{6\pi} \left( \gamma + \ln4\pi +\cdots \right)  (q_1^T C \sigma^{\alpha\beta} q_2)  \sigma_{\alpha\beta}  \gamma_{\mu}q_3, \quad \delta_{(c)} \left[ \eta^{(1)}_{\mu(12)3} \right]=0,
\end{align}
where $\gamma \simeq 0.5772$ is Euler-Mascheroni constant. Hence, one finds that the type $ \eta^{(1)}_{\mu(12)3} $ is renormalization covariant. 

The second type is defined as
 \begin{align}   
\eta^{(2)}_{\mu(12)3} &\equiv  (q_1^T C \gamma_{\mu}q_2) q_3.
\end{align}
For this current, the one-loop correction can be calculated as
\begin{align}  
\delta_{(a)} \left[ \eta^{(2)}_{\mu(12)3} \right]+\delta_{(b)} \left[ \eta^{(2)}_{\mu(12)3} \right]& =  \frac{\alpha_s}{12\pi} \left( \gamma + \ln4\pi +\cdots \right) \left(  2 (q_1^T C \gamma_{\mu} q_2)   q_3 - i (q_1^T C \gamma^\rho q_2)  \sigma_{\mu\rho}  q_3\right) ,\nonumber\\ 
\delta_{(c)} \left[ \eta^{(2)}_{\mu(12)3} \right] &= \frac{\alpha_s}{12\pi} \left( \gamma + \ln4\pi +\cdots \right)  (q_1^T C \gamma_{\mu} q_2)   q_3,
\end{align}
where  $(q^T C \gamma^\rho q)  \sigma_{\rho\mu}  q = i (q^T C \gamma_{\mu} q)   q$ in $q_1=q_2=q_3=q$ limit. The second type is renormalization covariant only when  $q_1=q_2=q_3$ condition is satisfied.  

A specific linear combination of the operators can have renormalization covariance. Using relations obtained through Fierz rearrangement,
\begin{align} 
  (u^T C \sigma^{\alpha\beta} u)  \sigma_{\alpha\beta}  \gamma_{\mu}d &=  8 (u^T C \gamma_{\mu} u) d - 2  (u^T C \gamma^{\rho} u) \gamma_{\rho} \gamma_{\mu}d,\\
  (u^T C \sigma^{\alpha\beta} d)  \sigma_{\alpha\beta}  \gamma_{\mu}u &=  2 (u^T C \gamma_{\mu} u) d +  (u^T C \gamma^{\rho} u) \gamma_{\rho} \gamma_{\mu}d,
\end{align}
one can obtain a renomalization covariant expression for the $\Delta^{+}$:
\begin{align} 
   \eta^{\Delta^{+}}_{\mu}\equiv  (u^T C \sigma^{\alpha\beta} u)  \sigma_{\alpha\beta}  \gamma_{\mu}d + 2 (u^T C \sigma^{\alpha\beta} d)  \sigma_{\alpha\beta}  \gamma_{\mu}u = 4 (u^T C \gamma_{\mu} u) d+ 8 (u^T C \gamma_{\mu} d) u, 
\end{align}
where the running off-diagonal terms are canceled. 

\section{Borel weighting scheme}\label{appenb}

\subsection{Borel sum rules in the vacuum}

Using analyticity  of the correlation function, the invariants can
be written in a dispersion relation:
\begin{align}
\Pi_i(q^2)&=\frac{1}{2\pi i}\int^\infty_{0} d s \frac{
\Delta\Pi_i(s)}{s-q^2} +P_n(q^2), \label{vaccord}
\end{align}
where $P_n(q^2)$ is a finite-order polynomial in $q^2$ which comes
from the integration on the circle of contour on complex plane and
the discontinuity
$\Delta\Pi_i(s)\equiv\lim_{\epsilon\rightarrow0^{+}}[\Pi_i(s+i\epsilon)-\Pi_i(s-i\epsilon)]=2i\textrm{Im}[\Pi_i(s+i\epsilon)]$
is defined on the positive real axis. All the possible physical
states are contained in this discontinuity. From phenomenological
considerations, the invariant can be assumed to have a pole and
continuum structure,
\begin{align}
\Delta\Pi_i(s)=\Delta\Pi_i^{\textrm{pole}}(s)+\theta(s-s_0)\Delta\Pi_i^{\textrm{OPE}}(s),\label{disca}
\end{align}
where $s_0$ represents the continuum threshold. To suppress  the
continuum contribution, weight function  $W(s)=e^{-s/M^2}$ can be
used as
\begin{align}
\mathcal{W}_M[\Pi_i(q^2)]=\frac{1}{2\pi i}\int^\infty_{0} d s~ e^{-s/M^2}
 \Delta\Pi_i(s).\label{borelv}
\end{align}
The corresponding differential operator $\mathcal{B}$ can be defined as
\begin{align}
\mathcal{B}[f(-q ^2) ]&\equiv \lim_{\substack{-q ^2,n\rightarrow\infty\\-q ^2/n=M^2}} \frac{(-q ^2)^{n+1}}{n!}\left(\frac{\partial}{\partial q ^2}\right)^n f(-q
^2).\label{boreltv}
\end{align}
By using this operator to Eq.~\eqref{vaccord}, one obtains following
relation:
\begin{align}
\mathcal{W}_M[\Pi_i(q^2)]&=\frac{1}{2\pi i}\int^\infty_{0} d s~ e^{-s/M^2}
 \Delta\Pi_i(s) =\mathcal{B}[\Pi_i(q^2)],\label{borelv1}
\end{align}
where following relation has been used:
\begin{align}
 \mathcal{B}\left[\frac{1}{s-q^2}\right] =e^{-s/M^2}.
\end{align}
The residues located after the continuum threshold with finite $s_0$
can be subtracted as
\begin{align}
\mathcal{W}_M^{\textrm{subt.}}[\Pi_i(q^2)]&= \frac{1}{2\pi i}\int^{s_0}_0 d s~ e^{-s/M^2}
 \Delta\Pi_i(s)=\mathcal{B}[\Pi_i(q^2)]_{\textrm{subt.}}.
\end{align}

Via simple integrations,
\begin{align}
\int_{s_0}^\infty ds e^{-s/M^2}&=M^2e^{-s_0 /M^2},\\
\int_{s_0}^\infty ds se^{-s/M^2}&=(M^2)^2e^{-s_0 /M^2}\left(s_0 /M^2+1\right),\\
\int_{s_0}^\infty ds s^2e^{-s/M^2}&=(M^2)^3e^{-s_0 /M^2}\left(s_0^{2}/2M^4+s_0 /M^2+1\right),
\end{align}
and the subtraction of continuum contribution for the OPE side can be summarized as
\begin{align}
E_0&\equiv1-e^{-s_0 /M^2},\\
E_1&\equiv1-e^{-s_0 /M^2}\left(s_0 /M^2+1\right),\\
E_2&\equiv1-e^{-s_0 /M^2}\left(s_0^{2}/2M^4+s_0 /M^2+1\right).
\end{align}
$E_n$ is multiplied to all $(M^2)^{n+1}$ terms in $\mathcal{B}[\Pi_i(q^2)]$.
This weighting scheme and subsequent spectral sum rules are known as Borel transformation and Borel sum rules.

\subsection{Borel sum rules in the medium}

Now one can extend this argument to the in-medium case. The OPE of the correlation function has been executed in the
$q_0^2 \rightarrow - \infty$,
$\vert\vec{q}\vert\rightarrow\textrm{fixed}$ limit and the
dispersion relation \eqref{dis-eo} has been written from Cauchy
relation in the complex energy $q_0$ space. As one can not expect
the symmetric spectral density assumed in the vacuum sum rules, the
contour (b) has been set to be different from the vacuum case, the
contour (a). Similarly, one can try following weighting in the
complex energy plane:
\begin{align}
\oint_{\textrm{contour}(b) }d\omega ~W(\omega)\Pi_i(\omega)&=0,\\
\int_{\substack{\textrm{circle}(b) \\ \vert \omega \vert =
\tilde{\omega}_0 }}d\omega~
W(\omega)\Pi_i(\omega)&=-\int_{-\tilde{\omega}_0}^{
\tilde{\omega}_0}d\omega
~W(\omega)\Delta\Pi_i(\omega).\label{weightb}
\end{align}
As the phenomenological structure \eqref{corrmed} is given with the
quasi-poles $E_q=\Sigma_v+\sqrt{\vec{q}^2+m_{\Delta}^{*2} }$ and
$\bar{E}_q=\Sigma_v-\sqrt{\vec{q}^2+m_{\Delta}^{*2} }$, the weight
function should suppress not only the OPE continuum but also the
quasi-anti-${\Delta}$ pole $\bar{E}_q$. $W(\omega)=
(\omega-\bar{E}_q)e^{-\omega^2/M^2}$ has been used for the in-medium
weight function for this purpose. This choice emphasizes the OPE
quasi-${\Delta}$ pole and suppresses the quasi-anti-${\Delta}$ pole and the
continuum. Moreover, in the zero density limit, the in-medium sum
rules obtained with $W(\omega)= (\omega-\bar{E}_q)e^{-\omega^2/M^2}$
can be reduced to the vacuum sum rules with an overall factor
$e^{-{\vec{q}^2/M^2}}$ as the discontinuity $\Delta\Pi_i(\omega)$ is
odd in the vacuum case. With the dispersion relations \eqref{dise}
and \eqref{diso}, the weighted sum of the residues can be written as
\begin{align}
\overline{\mathcal{W}}_M[\Pi_i(q_0,\vert\vec{q}\vert)]&=\frac{1}{2\pi
i}\int^\infty_{-\infty} d \omega~
(\omega-\bar{E}_q)e^{-\omega^2/M^2}
 \Delta\Pi_i(\omega^2,\vert\vec{q}\vert)\nonumber\\
 &=\frac{1}{2\pi
i} \left[\int^\infty_{-\infty} d \omega~ \omega^2
 e^{-\omega^2/M^2}\Delta\Pi^o_i(\omega^2,\vert\vec{q}\vert)-\bar{E}_q\int^\infty_{-\infty} d \omega~ e^{-\omega^2/M^2}
\Delta\Pi^e_i(\omega^2,\vert\vec{q}\vert)\right]\nonumber\\
&=\bar{\mathcal{B}}[\Pi^e_i(q^2,\vert\vec{q}\vert)]-\bar{E}_q\bar{\mathcal{B}}[\Pi^o_i(q^2,\vert\vec{q}\vert)],\label{borelm}
\end{align}
where the discontinuity $\Delta\Pi_i(\omega^2,\vert\vec{q}\vert)$
has been defined in Eq.~\eqref{discm} and the differential operator
$\bar{\mathcal{B}}$ is defined analogous with the the operator
\eqref{boreltv}:
\begin{align}
\bar{\mathcal{B}}[f(-q_0^2,\vert\vec{q}\vert)]\equiv
\lim_{\substack{-q_0^2,n\rightarrow\infty\\-q_0^2/n=M^2}}
\frac{(-q_0^2)^{n+1}}{n!}\left(\frac{\partial}{\partial
q_0^2}\right)^n f(-q_0^2,\vert\vec{q}\vert).\label{boreltm}
\end{align}
The subtraction of the residues after $s_0^{*}=\omega_0^2-\vec{q}^2$
can be summarized as Eqs.~\eqref{contsub0}-\eqref{contsub2},
analogous with the subtraction scheme of the vacuum case.

The main difference from the vacuum Borel sum rules is the
appearance of the quasi-anti-pole $\bar{E}_q$:
\begin{align}
\overline{\mathcal{W}}_M^{\textrm{subt.}}[\Pi_i(q_0,\vert\vec{q}\vert)]=\left[\bar{\mathcal{B}}[\Pi^e_i(q^2,\vert\vec{q}\vert)]
-\bar{E}_q\bar{\mathcal{B}}[\Pi^o_i(q^2,\vert\vec{q}\vert)]\right]_{\textrm{subt.}}.
\end{align}
The information from the residue of the quasi-baryon pole has been
isolated following this scheme.

\end{document}